\documentclass[11pt]{article}
\usepackage{amsmath}
\usepackage{graphicx}
\usepackage{enumerate}
\usepackage{amsfonts}
\usepackage{url} % not crucial - just used below for the URL 

\usepackage[margin=10mm]{caption}
\usepackage{geometry}
\usepackage{mathptmx}
\usepackage{graphics,epsfig,latexsym,amssymb,amsbsy,amsmath}
\usepackage{amsthm}
\usepackage[margin=15mm]{caption}

\renewcommand{\baselinestretch}{1.2}
\newcommand{\bX}{{\bf X}}

\newcommand{\bA}{{\bf A}}
\newcommand{\bD}{{\bf D}}
\newcommand{\bx}{{\bf x}}
\newcommand{\btheta}{\mbox{\protect\boldmath $\theta$}}

\newcommand{\bbeta}{\mbox{\protect\boldmath $\beta$}}
\newcommand{\bLambda}{\mbox{\protect\boldmath $\Lambda$}}
\newcommand{\rP}{\mbox{P}}

\newcommand{\one}{1\!\!{\rm I}}

\newcommand{\etab}{\mbox{\protect\boldmath $\eta$}}

\newcommand{\rmd}{\mathrm{d}}

\begin{document}

\def\spacingset#1{\renewcommand{\baselinestretch}%
{#1}\small\normalsize} \spacingset{1.2}

{
\title{Bayesian Federated Inference for estimating Statistical Models based on Non-shared Multicenter Data sets}
  \author{Marianne A Jonker \thanks{corresponding author: marianne.jonker@radboudumc.nl}\hspace{.2cm}\\
    Research Institute for Medical Innovation, Science department IQ Health, Section Biostatistics, \\
    Radboudumc, Nijmegen, The Netherlands\\
    and \\
    Hassan Pazira \\
    Research Institute for Medical Innovation, Science department IQ Health, Section Biostatistics, \\
    Radboudumc, Nijmegen, The Netherlands\\
    and \\
Anthony CC Coolen\\
    Donders Institute, Faculty of Science, Radboud University, \\
    Nijmegen, The Netherlands\\
    Saddle Point Science Europe, Mercator Science Park, \\
    Nijmegen, The Netherlands
    } 
}  

\maketitle

\abstract{Identifying predictive factors for an outcome of interest via a multivariable analysis is often difficult when the data set is small. Combining data from different medical centers into a single (larger) database would alleviate this problem, but is in practice challenging due to regulatory and logistic problems. Federated Learning (FL) is a machine learning approach that aims to construct from local inferences in separate data centers what would have been inferred had the data sets been merged. It seeks to harvest the statistical power of larger data sets without actually creating them. The FL strategy is not always efficient and precise. Therefore, in this paper we refine and implement an alternative Bayesian Federated Inference (BFI) framework for multicenter data with the same aim as FL. The BFI framework is designed to cope with small data sets by inferring locally not only the optimal parameter values, but also additional features of the posterior parameter distribution, capturing information beyond what is used in FL. BFI has the additional benefit that a single inference cycle across the centers is sufficient, whereas FL needs multiple cycles. We quantify the performance of the proposed methodology on simulated and real life data.\\

\bigskip

\noindent%
}

\maketitle

\section{Introduction}
Medical data are only occasionally easily shared for research by those who hold them (companies, hospitals, research centers, universities, and consortia). The lack of sharing hampers more accurate estimation and inference. 
Often, in existing data sets the number of patients is small compared to the number of covariates available for prediction, 
which leads to misidentification of prognostic factors, overfitting of the statistical model and finally to inaccurate predictions for future patients. This is often the case for rare diseases, for which only small data sets are typically available for research.
To ease the problem, we must either create more effective mechanisms and incentives for data sharing between institutions or focus on technology to integrate individual analysis outcomes obtained on localized data sets. If we go for the second route, the scientific problem is how to extract and combine information from different analyses on different non-overlapping and exhausting subsets to obtain the estimates that would have been found had the subsets been combined into a single data set. 

\begin{figure}
\centering
\centering
\includegraphics[scale=0.49]{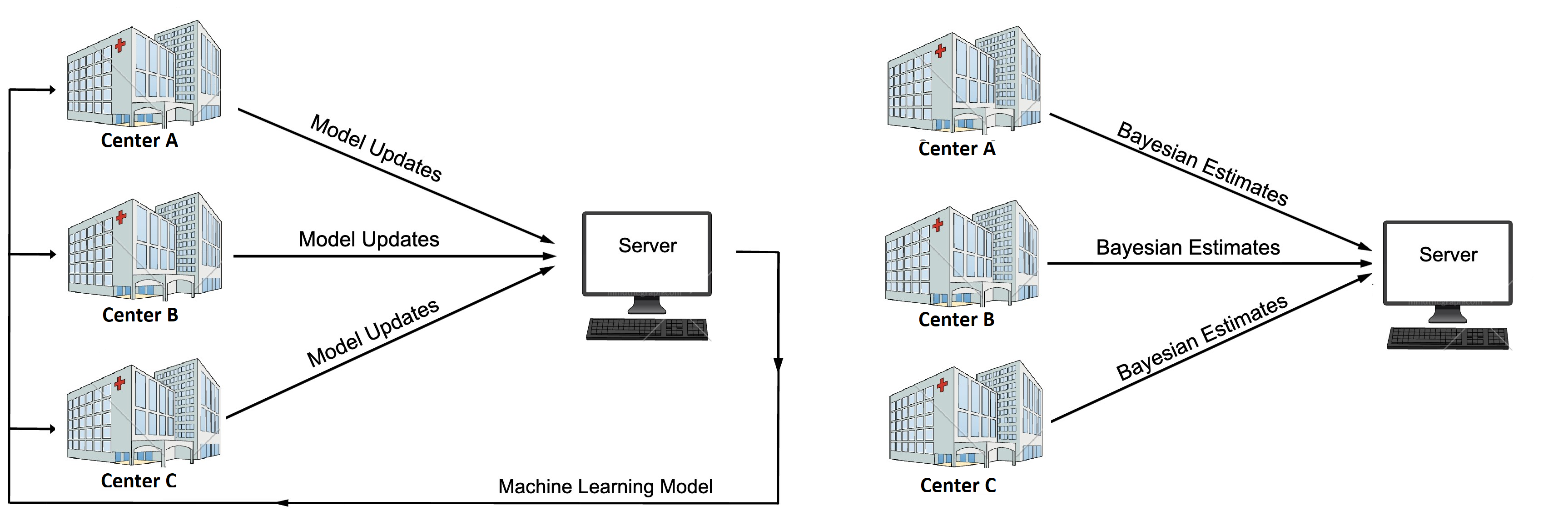}
 \caption{Left: Visualization of an iterative FL procedure. In the local centers local loss functions are optimized based on local data. The results are aggregated at a central server. Next, the estimates are updated based on the local data and the aggregated information. This procedure is repeated until convergence. Right: Visualization of the BFI procedure proposed in this paper. In all local centers a Bayesian analysis is performed and the results are sent to a server where the results of the local centers are aggregated. The word “Learning” in the name “Federated Learning” refers to the repeated cycling mechanism around different centers. In contrast, for the proposed method one cycle is enough; inference is performed only once. We therefore chose to name the proposed method Bayesian Federated Inference and not Bayesian Federated Learning.
}
\label{fig:FL}
\end{figure}

Federated Learning (FL) is a %an online (i.e., non-batch) 
machine learning approach, developed several years ago in the context of interacting mobile devices\cite{McMahan}, %(McMahan et al., 2017\nocite{McMahan}), 
in which local centers use local data for training machine learning systems on site by optimizing a pre-specified loss function, and only the estimated parameters of the trained systems are sent out for integration at a central server. The data, in contrast, stay at their owners' institutions. This procedure is `cycled' around the centers iteratively; all centers update their estimates %in an online fashion 
until a convergence criterion is met (see Figure \ref{fig:FL}, left plot). Typically, deep learning neural networks are employed, where the circulated parameters are strengths of interactions between nodes and activation thresholds of nodes. 
While FL performs excellently in image analysis \cite{Rieke, Sheller, Gafni}, %(see e.g., Rieke et al., 2020; Sheller et al., 2020; Gafni et al., 2021\nocite{Rieke, Sheller, Gafni}), 
it also has limitations in application. First, in its standard form FL still requires relatively large data sets (given the large numbers of deep learning parameters). Second, FL does not generate rigorous error bars, since only the most probable parameter values are iterated (as opposed to posterior distributions). Third, it is not clear that this process will always converge to a satisfactory final state, since the latent heterogeneity across centers may generate prohibitively conflicting gradients for the model parameters in the maximization procedure, leading to an end result that works for none of the local data sets.  

In recent years much progress has been made in FL (e.g., improved optimization in each center, aggregation of the local models at the central server and dealing with heterogeneity of the local populations \cite{Li2020, Chen, Zhu} %(Li et al., 2020; Chen and Chao, 2020; Zhu et al., 2021\nocite{Li2020b, Chen, Zhu}). % [Li et al., 2020b, Chen and Chao, 2020, Zhu et al.,2021]
). 
An overview of the recent developments including references is given in Liu et al.\cite{Liu}. Multiple researchers have proposed FL in a Bayesian setting for deep learning models\cite{Maddox, Al-Shedivat}. 
%(e.g., Maddox et al., 2019; Al-Shedivat et al., 2020\nocite{Maddox, Al-Shedivat}): 
In each local center the posterior distribution is estimated and communicated to the central server, where aggregation takes place. However, this Bayesian procedure turns out to be challenging, especially for deep learning models in which the parameter dimensionality is high. Proposals to address this include approximating the estimated local and global posterior distribution with MCMC \cite{Zhang2019,Izmailov},
%(Zhang et al., 2019; Izmailov et al., 2021)\nocite{Zhang2019, Izmailov}, 
variational inference\cite{Zhang2018},
%(Zhang et al., 2018)\nocite{Zhang2018}), 
or the use of  multivariate Gaussian distributions with a Laplace approximation\cite{Maddox,Al-Shedivat}. 
%(Maddox et al.\ 2019; Al-Shedivat et al.\ 2020\nocite{Maddox,Al-Shedivat}). 
For a linear model (as an example)
Al-Shedivat et al.\ \cite{Al-Shedivat} approximated the local posterior by a multivariate Gaussian. Under the assumption of a flat prior for the model parameters in the local centers, the global posterior distribution in the central server is multivariate Gaussian as well, and estimates of its parameters can be computed from the local sample means and covariance matrices. Although this seems to be an interesting idea on paper, the authors directly note that this way of estimating the global posterior distribution is not feasible in a general setting where models are neural networks with millions of parameters, as it requires estimating and computing the inverse of an $d\times d$ matrix, where $d$ is the number of parameters. 

In the field of statistical modelling (e.g., via generalized linear models) the challenges in FL are not about the computational burden due to the high dimensionality of the model parameters,
but rather about the complexities of (medical) data, such as heterogeneity of populations, random or structural missingness of covariates, presence of confounding factors, overfitting,  and the interpretation of the parameters in the estimated models. In this paper a Bayesian Federated Inference (BFI) framework is proposed and developed for arbitrary generalized linear regression models, in which the challenges described above are addressed. More specifically, the  posterior distributions in the local centers and for the fictive combined data are approximated by multivariate Gaussian distributions around the maximum a posteriori (MAP) estimate. By choosing either a multivariate Gaussian or uninformative distribution for the prior distribution, the outcome of the inference on the combined set can be expressed directly in terms of the outcomes of the separate inferences on the local data sets. With the proposed methodology, there is no need to do inference on the full data set, as its inference results can be computed {\em a posteriori} from the inference results in the subsets in only one round, in contrast to traditional FL where very many iterative inference cycles across the centers are needed (see plot on the right in Figure \ref{fig:FL}). The proposed BFI method is also accurate when the sample sizes are small compared to the dimension of the parameter space, by choosing a more informative prior to overcome overfitting of the model.

This paper is organized as follows. In Section \ref{sec:BFI} the framework for the BFI strategy is explained in a general setting. Next, in Section \ref{Sec:GLM} we focus on BFI for generalized linear models. Here we also address population heterogeneity and missing covariates. Simulation studies based on real life data have been performed to quantify the efficiency in statistical inference if the data are only locally available for analysis, and the impact of complexities related to heterogeneity and (relative) small sample sizes in the local centers (Section \ref{sec:simstudies}). The paper ends with a discussion in Section \ref{sec:Discussion}.

\section{Bayesian Federated Inference (BFI)}
\label{sec:BFI}

\subsection{Problem definition and setting}
Suppose data from $L$ different centers (e.g., hospitals) are available. Let random variable $Y_{\ell i}$ be the $i^{th}$ observation from the $\ell^{th}$ center, where $i=1,\ldots,n_\ell$ with $n_\ell$ defined as the number of observations in center $\ell$. We assume that $Y_{\ell i}, i=1,\ldots,n_\ell,\ell=1,\ldots, L,$ are stochastically independent and identically distributed. Specifically, we assume that $Y_{\ell i}|\btheta$ is distributed with parametric density function %\footnote{We use the letter $p$ for any density. From the arguments it follows which density is meant.} 
$y|\btheta \to p(y|\btheta)$ that is known up to the parameter $\btheta$, which itself has a density function $\btheta\to p(\btheta)$. The realisation of $Y_{\ell i}$, i.e., the actual observation, is denoted as $y_{\ell i}$. The $L$ data sets $\bD_1,\ldots, \bD_L$ and their union are defined as:  
\begin{eqnarray}
\bD_\ell=\{y_{\ell 1},\ldots,y_{\ell n_\ell}\},~~~~~~~~n_\ell=|\bD_\ell|,~~~~~~~~\bD=\bigcup_{\ell=1}^L \bD_\ell,~~~~~~~~ n=|\bD|=\sum_{\ell=1}^L n_\ell. \nonumber
\end{eqnarray}
We imagine the scenario where the data from the $L$ sets are prohibited from being combined into one data set $\bD$; inference can only be performed for the $L$ sets separately, and only the inference results can be combined. In the next subsection we aim to express the outcome of inference on the combined set $\bD$ in terms of the outcomes of the $L$ separate inferences on the constituent sets $\bD_\ell$. 

\subsection{Formulae for Bayesian subset inference}
\label{sub:formulae}

In a Bayesian analysis based on any data set $\bD=\{y_1,\ldots,y_n\}$ from a conditional density function $y|\btheta\to p(y|\btheta)$ with prior density $\btheta\to p(\btheta)$, the posterior density equals:
\begin{eqnarray}
p(\btheta |\bD)\;=\;\frac{p(\bD|\btheta)p(\btheta)}{Z(\bD)},~~~~~~~~~~Z(\bD)\;=\;\int\!p(\bD|\btheta)p(\btheta)\rmd\btheta,~~~~~~~~~~~p(\bD|\btheta)=\prod_{i=1}^{n}p(y_i|\btheta),
\label{Eq:post}
\end{eqnarray}
where $Z(\bD)$ denotes the normalizing constant of the posterior distribution.
The maximum a posteriori (MAP) estimator\cite{MacKay} for $\btheta$ is obtained by maximizing the function $\btheta|\bD\to p(\btheta|\bD)$ with respect to $\btheta$.
Since the complete data set $\bD$ is the union of $L$ data sets from different centers, we can write the  posterior density function as
\begin{eqnarray}
p(\btheta|\bD)\;=\; \frac{p(\btheta)}{Z(\bD)}\prod_{\ell=1}^L \prod_{i=1}^{n_\ell}p(y_{\ell i}|\btheta).
\label{Ex:post2}
\end{eqnarray}
The expressions in (\ref{Eq:post}) hold for any data set, so they apply also to the sets $\bD_\ell,  \ell=1,\ldots,L$. By rewriting the expression in (\ref{Ex:post2}) with $\bD$ replaced by $\bD_\ell$, and with the subset priors denoted by $p_\ell(\btheta)$, we find that  
\begin{eqnarray}
\prod_{i=1}^{n_\ell}p(y_{\ell i}|\btheta)
 &\!=\!& \frac{Z_\ell(\bD_\ell)}{p_\ell(\btheta)} \; p_\ell(\btheta|\bD_\ell), ~~~~~~~~
 Z_\ell(\bD_\ell)\;=\;\int\!p(\bD_\ell|\btheta)p_\ell(\btheta)\rmd\btheta. \nonumber
\end{eqnarray}
By substituting this expression into (\ref{Ex:post2}), we obtain:
\begin{eqnarray}
p\big(\btheta|\bD\big)&\!=\!& 
\frac{1}{C}\;\bigg( \frac{p(\btheta)}{\prod_{\ell=1}^L p_\ell(\btheta)}\bigg)
\prod_{\ell=1}^L p_\ell(\btheta|\bD_\ell),~~~~\mbox{  with  }~~~~ C=\frac{Z(\bD)}{\prod_{\ell=1}^L Z_\ell(\bD_\ell)}.
\label{eq:link}
\end{eqnarray}
The constant $C$ can always be recovered via normalization. 
Hence we can in a relatively simple way express the posterior parameter density $p(\btheta|\bD)$ in terms of the $L$ local densities $p_\ell(\btheta|\bD_\ell)$. 

The next question is under which conditions  we can obtain accurate (possibly approximate) expressions for the MAP estimator $\widehat{\btheta}$ of $\btheta$ (and its accuracy), which is based on data set $\bD$, from the $L$ MAP estimators $\widehat{\btheta}_\ell, \ell=1,\ldots,L$ (and their accuracy), which are solely inferred from the data subsets $\bD_\ell$. This is the topic for the next  subsection.

\subsection{Bayesian Federated Inference}

If the distribution of the data comes from an exponential family, BFI is straightforward; the MAP estimator based on the data $\bD$ can be obtained directly from summary statistics in the $L$ subsets. Once models become more complex approximating the  posterior density $p(\btheta|\bD)$ by a multivariate Gaussian density may be needed. The Bernstein-von Mises theorem states that under certain regularity conditions and for sufficiently large sample size $n$, the posterior distribution can be approximated well by a multivariate Gaussian distribution centered at the maximum likelihood estimator of $\btheta$.\cite{Vaart} In practice, the sample sizes in the centers are finite and may even be small compared to the number of covariates. We, therefore, center around the MAP estimator. We approximate the logarithm of the posterior density (given the combined data set $\bD$)
\begin{eqnarray}
\log \big\{ p\big(\btheta|\bD\big) \big\} \;=\; \log \big\{ p\big(\btheta\big)\big\} + \sum_{\ell=1}^L \sum_{i=1}^{n_\ell}\log \big\{ p\big(y_{\ell i}|\btheta\big)\big\} -\log \big\{Z\big(\bD\big)\big\} \nonumber
\end{eqnarray}
by a Taylor expansion  up to a quadratic order in $\btheta$ around its MAP parameter estimator $\widehat{\btheta}$ which is based on the combined data set $\bD$: 
\begin{eqnarray}
\log \big\{ p\big(\btheta|\bD\big) \big\} \;=\; \log \big\{p\big(\widehat{\btheta}|\bD\big)\big\} -\tfrac{1}{2}\big(\btheta-\widehat{\btheta}\big)^t \widehat{\bA}\big(\btheta-\widehat{\btheta}\big) + O_p\big(\|\btheta-\widehat{\btheta}\|^3\big),
\label{Taylor}
\end{eqnarray}
where $\widehat{\bA}$ is equal to minus the curvature matrix of $\log \{p(\btheta|\bD)\}$ in $\widehat{\btheta}$, i.e., the matrix of minus second order derivatives  with respect to $\btheta$, evaluated at $\widehat{\btheta}$. The last term in (\ref{Taylor}) is  $O_p(\|\btheta-\widehat{\btheta}\|^3)=\|\btheta-\widehat{\btheta}\|^3 O_p(1)$ which is equal to $\|\btheta-\widehat{\btheta}\|^3$ times the term $O_p(1)$, where $O_p(1)$ represents a term that is bounded in probability for the sample size to infinity\cite{Vaart}. Note that there is no linear term in the Taylor expansion (\ref{Taylor}), because the gradient of $\log \{p(\btheta|\bD)\}$ at $\widehat{\btheta}$ equals zero by definition (as $\widehat{\btheta}$ is the MAP estimator). From (\ref{Taylor}) it follows that  
\begin{eqnarray}
p\big(\btheta|\bD\big) \;\propto \; \exp\big\{-\tfrac{1}{2}\big(\btheta-\widehat{\btheta}\big)^t \widehat{\bA}\big(\btheta-\widehat{\btheta}\big) + O_p\big(\|\btheta-\widehat{\btheta}\|^3\big)\big\}, \nonumber
\end{eqnarray}
(seen as a function of $\btheta$).
For $\btheta$ in a sufficiently small neighbourhood of $\widehat{\btheta}$, the term $O_p(\|\btheta-\widehat{\btheta}\|^3)$ will be close to zero compared to the quadratic term. Then, the posterior density in a small neighbourhood of $\widehat{\btheta}$ can be approximated by a Gaussian density: 
\begin{eqnarray}
p\big(\btheta|\bD\big) \;\approx\; \bigg(\frac{{\rm det} (\widehat{\bA})}{(2\pi)^d}\bigg)^{\frac{1}{2}}
\exp\big\{-\tfrac{1}{2}\big(\btheta-\widehat{\btheta}\big)^t\widehat{\bA}\big(\btheta-\widehat{\btheta}\big)\big\},
\label{exp:Gaussian}
\end{eqnarray}
where $d={\rm dim}(\btheta)$ and ${\rm det} (\widehat{\bA})$ denotes the determinant of the matrix $\widehat{\bA}$. In case of high dimensional models, this approximation may not be sufficiently accurate. A higher order Taylor expansion might be needed. This is discussed in Section \ref{sec:Discussion}.
In a similar way, each of the $L$ posterior distributions for the subsets $\ell=1\ldots L$ is approximately Gaussian around $\widehat{\btheta}_\ell$ (the MAP estimator based on data set $\bD_\ell$) and with covariance matrix equal to the inverse of  $\widehat{\bA}_\ell$. 
By substituting expression (\ref{exp:Gaussian}) for $p(\btheta|\bD)$, and the equivalent expressions for $p_\ell(\btheta|\bD_\ell),\ell=1,\ldots,L$, into relation (\ref{eq:link}), we obtain:
\begin{eqnarray*}
\bigg(\frac{{\rm det} (\widehat{\bA})}{(2\pi)^d}\bigg)^{\frac{1}{2}}
\exp\big\{-\tfrac{1}{2}\big(\btheta-\widehat{\btheta}\big)^t \widehat{\bA} \big(\btheta-\widehat{\btheta}\big)\big\} \; \approx   \;\frac{\prod_{\ell=1}^L Z_\ell(D_\ell)}{Z(D)}~
\frac{p(\btheta)}{\prod_{\ell=1}^L p_\ell(\btheta)}
\prod_{\ell=1}^L  \bigg(\frac{{\rm det} (\widehat{\bA}_\ell)}{(2\pi)^d}\bigg)^{\frac{1}{2}}\exp\big\{-\tfrac{1}{2}\big(\btheta-\widehat{\btheta}_\ell\big)^t \widehat{\bA}_\ell\big(\btheta-\widehat{\btheta}_\ell\big)\big\}. \qquad
%\label{eq:setequal}
\end{eqnarray*} 

\bigskip

\noindent
If the prior densities $\btheta\to p(\btheta)$ and $\btheta\to p_\ell(\btheta)$  are chosen to be Gaussian, with mean zero and covariance matrices $\bLambda^{-1}$ and $\bLambda_\ell^{-1}$, i.e., 
\begin{align*}
p(\btheta) = \bigg(\frac{\det (\bLambda)}{(2\pi)^d}\bigg)^{\frac{1}{2}}\exp\big(-\tfrac{1}{2}\btheta^t \bLambda \btheta\big) ~~~~~~~\mbox{ and } ~~~~~~~~ p_\ell(\btheta) = \bigg(\frac{\det (\bLambda_\ell)}{(2\pi)^d}\bigg)^{\frac{1}{2}} \exp\big(-\tfrac{1}{2}\btheta^t \bLambda_\ell \btheta\big),
\end{align*}
the above equation becomes
\vspace{10pt}
\begin{eqnarray*}
\lefteqn{\bigg(\frac{{\rm det} (\widehat{\bA})}{(2\pi)^d}\bigg)^{\frac{1}{2}}
\exp\big\{-\tfrac{1}{2}\big(\btheta-\widehat{\btheta}\big)^t \widehat{\bA} \big(\btheta-\widehat{\btheta}\big)\big\} } \nonumber \\[6pt]
&& \;\approx \;  \; \frac{\prod_{\ell=1}^L Z_\ell(\bD_\ell)}{(2\pi)^{\frac{1}{2}d}Z(\bD)}
\Bigg(
\frac{\det (\bLambda)\prod_{\ell=1}^L 
{\rm det} (\widehat{\bA}_\ell)}{\prod_{\ell=1}^L \det (\bLambda_\ell)}\Bigg)^{\!\frac{1}{2}} 
\; \; \exp\Big\{-\tfrac{1}{2}\btheta^t \big(\bLambda-\sum_{\ell=1}^L  \bLambda_\ell\big) \btheta\Big\}
\prod_{\ell=1}^L 
\exp\Big\{-\tfrac{1}{2}\big(\btheta-\widehat{\btheta}_\ell\big)^t \widehat{\bA}_\ell\big(\btheta-\widehat{\btheta}_\ell\big)\Big\}.
\end{eqnarray*}
Taking the logarithm on both sides yields, when viewed as a function of $\btheta$:
\begin{eqnarray}
\big(\btheta-\widehat{\btheta}\big)^t \widehat{\bA} \big(\btheta-\widehat{\btheta}\big)
&\approx&
\btheta^{t} \Big(\bLambda-\sum_{\ell=1}^L  \bLambda_\ell\Big) \btheta
+\sum_{\ell=1}^L \big(\btheta-\widehat{\btheta}_\ell\big)^t \widehat{\bA}_\ell\big(\btheta-\widehat{\btheta}_\ell\big)+B,
\label{eq:log-equality}
\end{eqnarray}
with $B$ representing a term that is a function of the data, but does not depend on $\btheta$. The expressions on both sides are now considered as functions of $\btheta$. Both sides are quadratic functions of $\btheta$, and must have the same expansion coefficients for the linear and quadratic terms in $\btheta$. This implies that we can find estimators $\widehat{\btheta}$ and $\widehat{\bA}$ by solving the equations:
\begin{eqnarray*}
{\rm quadratic~terms:}&& \sum_{j,k=1}^d\theta_j\theta_k\Big(
\widehat{A}_{jk}
-\Lambda_{jk}+ \sum_{\ell=1}^L\Lambda_{\ell,jk}-\sum_{\ell=1}^L  \widehat{A}_{\ell,jk} \Big)=0   ~~~~~~~~~~~~ \forall \btheta
\\
{\rm linear~terms:}&& \sum_{j=1}^d \theta_j\Big(\widehat{\bA}\; \widehat{\btheta}-\sum_{\ell=1}^L \widehat{\bA}_\ell\widehat{\btheta}_\ell\Big)_j=0 ~~~~~~~~~~~~ \forall \btheta.
\end{eqnarray*}
Solving these equations by using the fact that they must hold for any value of $\btheta$ (in a neighbourhood of $\widehat{\btheta}$), yields:  
\begin{eqnarray}
\widehat{\bA}\;\approx\; \sum_{\ell=1}^L \widehat{\bA}_\ell+\bLambda-\sum_{\ell=1}^L \bLambda_\ell,
~~~~~~~~~~~~~
\widehat{\btheta} \;\approx\; \Big(\sum_{\ell=1}^L \widehat{\bA}_\ell+\bLambda-\sum_{\ell=1}^L \bLambda_\ell\Big)^{-1}\sum_{\ell=1}^L \widehat{\bA}_\ell\widehat{\btheta}_\ell \;=\; \big(\widehat{\bA}\big)^{-1} \sum_{\ell=1}^L \widehat{\bA}_\ell\widehat{\btheta}_\ell.
\label{eq:recombination_formulae}
\end{eqnarray}
The expression for $\widehat{\bA}$ follows from the equality with the quadratic terms in $\btheta$, and the expression for $\widehat{\btheta}$ from the equality with the linear terms. There is no need to do inference on the combined data set $\bD$ to find (an approximation of) the estimator $(\widehat{\btheta},\widehat{\bA})$ from the combined set; we can compute approximations {\em a posteriori} from the inference results on the subsets. 
The MAP estimators $\widehat{\btheta}_\ell$ and the matrices $\widehat{\bA}_\ell$ are found analytically or numerically, depending on the form of the likelihood functions in the local centers. 
%For any choice of the positive definite matrix $\bLambda_\ell$, the matrix $\widehat{\bA}_\ell$ is semi-positive definite by definition. This will also hold for the sum: $\sum_{\ell=1}^L \widehat{\bA}_\ell$. 
The matrix $\widehat{\bA}$ is positive definite by definition, so its approximation in (\ref{eq:recombination_formulae}) as well (if the approximation is sufficiently accurate) and, thus, is invertible.

By its definition as minus the second derivatives of $\log \{p(\btheta|\bD_\ell)\}$ at $\widehat{\btheta}_\ell$, $\widehat{\bA}_\ell$ equals
\begin{eqnarray}
\widehat {\bA}_\ell \;=\; \bLambda_\ell - \bigg(\frac{\partial^2}{\partial \btheta_j \partial \btheta_k } \sum_{i=1}^{n_\ell} \log \big\{p(y_{\ell i}|\btheta)\big\}\bigg)_{j,k}\Big|_{\btheta=\widehat{\btheta}_\ell}. \nonumber
\end{eqnarray}
%For $\widehat{\btheta}_\ell$ close enough to the maximum likelihood estimator of $\btheta$ minus the Hessian matrix of the log likelihood function in $\widehat{\btheta}_\ell$ will be semi-negative definite. 
Inserting the expression of $\widehat{\bA}_\ell$ into the expression for $\widehat{\bA}$ in (\ref{eq:recombination_formulae}), gives  
\begin{eqnarray*}
\widehat{\bA} \;\approx\; \bLambda - \sum_{\ell=1}^L \Bigg\{\bigg(\frac{\partial^2}{\partial \btheta_j \partial \btheta_k } \sum_{i=1}^{n_\ell} \log p(y_{\ell i}|\btheta)\bigg)_{j,k}\Big|_{\btheta=\widehat{\btheta}_\ell} \Bigg\}, 
\end{eqnarray*}
where the dependence on $\bLambda_\ell$ is via the MAP estimator $\widehat{\btheta}_\ell$ only. If  $\widehat{\btheta}_1 = \ldots = \widehat{\btheta}_L$,  the expression for $\widehat{\bA}$ equals minus the second derivative of $\log \{p(\btheta | \bD)\}$, the log posterior distribution after combining all data.

Even if one only wishes to do MAP inference, one would still need the matrix $\widehat{\bA}$ to compute error bars or regions for the MAP estimator $\widehat{\btheta}$. More specifically, for the $k^{th}$ element of $\btheta$ its approximate $(1-2\alpha) 100\%$ credible interval equals
$\widehat{\btheta}_k \pm \xi_\alpha \; \big(\widehat{\bA}^{-1}\big)_{k,k}^{1/2},$
for $\xi_\alpha$ the upper $\alpha$-quantile of the standard Gaussian distribution. Hypothesis testing is also straightforward by the asymptotic normality of $\widehat{\btheta}$.

The formulae in (\ref{eq:recombination_formulae}) do not say anything about the plausibility of the subsets describing similar subpopulations. However, once we have computed the estimates $(\widehat{\btheta},\widehat{\bA})$ we should find that $\widehat{\btheta}$ is compatible with each `local' estimate $\widehat{\btheta}_\ell$, given the error bars coded in the matrices $\widehat{\bA}$ and $\widehat{\bA}_\ell$. 
One way to address this is to consider the coordinate wise approximate $(1-2\alpha) 100\%$ credible intervals for the difference  between the true $\btheta$-value in all subsets except subset $\ell$ and the true parameter value in subset $\ell$:
\begin{align}
\big(\widehat{\btheta}_{-\ell}-\widehat{\btheta}_{\ell}\big)_k \; \pm \; \xi_\alpha \; \sqrt{\Big\{\big(\widehat{\bA}_{-\ell}\big)^{-1}+\big(\widehat{\bA}_\ell\big)^{-1}\Big\}_{kk}},   \nonumber
\end{align} 
where subscript $-\ell$ indicates that the BFI estimator excluded the estimator from subset $\ell$.

\subsection{Nuisance parameters differ across centers}
\label{subsec:nuisance}

Suppose that the vector $\btheta$ can be decomposed as $\btheta^t=(\btheta_a^t, \btheta_b^t)$, where $\btheta_a$ denotes the core parameters of interest of dimension $d_1$, and $\btheta_b$ the nuisance parameters of dimension $d_2$, with $d_1+d_2=d$. The parameter vector of interest $\btheta_a$ is assumed to be equal in the $L$ different sub-populations, but  $\btheta_b$ may vary across the sub-populations. Let $\btheta_{b,\ell}$ be the vector of nuisance parameters in population $\ell$. Then, the parameter vector in the model for the combined data set is equal to $\btheta^t=(\btheta_a^t;\btheta_{b,1}^t, \ldots, \btheta_{b,L}^t)$, and is of dimension $d^\prime = d_1+Ld_2$. 
Let $\widehat{\btheta}_{b,\ell}$ be the MAP estimate of $\btheta_{b,\ell}$ based solely on data in the $\ell^{th}$ subset, and $\widetilde{\btheta}_{b,\ell}$ be the MAP estimate based on the full data set. We choose simple priors of the form $p(\btheta)=p(\btheta_a)\prod_{\ell=1}^L p(\btheta_{b,\ell})$ for the combined data set, and $p_\ell(\btheta_a,\btheta_{b,\ell})=p_\ell(\btheta_a)p_\ell(\btheta_{b,\ell})$ in data subset $\ell$. We then have 
\begin{eqnarray}
\log \big\{p(\btheta|\bD)\big\} &=& \log \big\{p(\btheta_a)\big\} + \sum_{\ell=1}^L\sum_{i=1}^{n_\ell} \log\big\{p(y_{\ell i}|\btheta_a,\btheta_{b,\ell})\big\} + 
\sum_{\ell=1}^L \log \big\{p(\btheta_{b,\ell})\big\} -\log \big\{Z(\bD)\big\},
 \nonumber \\
\log \big\{p_\ell(\btheta_a,\btheta_{b,\ell}|\bD_\ell)\big\} &=&  \log \big\{p_\ell(\btheta_a)\big\} +\sum_{i=1}^{n_\ell} \log \big\{p(y_{\ell i}|\btheta_a,\btheta_{b,\ell})\big\} + \log \big\{p_\ell(\btheta_{b,\ell})\big\}  -\log \big\{Z_\ell(\bD_\ell)\big\}, \nonumber
\end{eqnarray}
and hence, like in (\ref{eq:link}),
\begin{eqnarray}
\hspace*{-3mm}
\log \big\{p(\btheta|\bD)\big\} &=& \sum_{\ell=1}^L \log \big\{p_\ell(\btheta_a,\btheta_{b,\ell}|\bD_\ell)\big\} + \log\Bigg(\frac{p(\btheta_a)}{\prod_{\ell=1}^L p_\ell(\btheta_a)}\Bigg) \;+ \; \log\Bigg(\frac{\prod_{\ell=1}^L p(\btheta_{b,\ell})}{\prod_{\ell=1}^L p_\ell(\btheta_{b,\ell})}\Bigg)-\log\Bigg(\frac{Z(\bD)}{\prod_{\ell=1}^L Z_\ell(\bD_\ell)}\Bigg).~~~~
\label{eq:linkyonly}
\end{eqnarray}
The quadratic Taylor approximations of the log posterior densities $\log \{p\big(\btheta|\bD\big)\}$ and $\log \{p_\ell\big(\btheta_a,\btheta_{b,\ell}|\bD_\ell\big)\}$ for $\btheta$ in a neighbourhood of $\widehat{\btheta}=(\widehat{\btheta}_a,\widetilde{\btheta}_{b,1},\ldots,\widetilde{\btheta}_{b,L})$ and $(\widehat{\btheta}_{a,\ell},\widehat{\btheta}_{b,\ell})$ respectively, are now
\begin{eqnarray}
\log \big\{p\big(\btheta|\bD\big)\big\} &\approx& \log \big\{p\big(\widehat{\btheta}|\bD\big)\big\} -\tfrac{1}{2}\big(\btheta_a-\widehat{\btheta}_a\big)^t \widehat{\bA}_a\big(\btheta_a-\widehat{\btheta}_a\big) \nonumber\\
&& -\;\tfrac{1}{2}\sum_{\ell=1}^L \big(\btheta_{b,\ell}-\widetilde{\btheta}_{b,\ell}\big)^t \widetilde{\bA}_{b,\ell}\big(\btheta_{b,\ell}-\widetilde{\btheta}_{b,\ell}\big)-\big(\btheta_a-\widehat{\btheta}_a\big)^t\sum_{\ell=1}^L \widetilde{\bA}_{ab,\ell}\big(\btheta_{b,\ell}-\widetilde{\btheta}_{b,\ell}\big),
\nonumber \\[1mm]
\log \big\{p_\ell\big(\btheta_{a},\btheta_{b,\ell}|\bD_\ell\big)\big\} &\approx& \log \big\{ p_\ell\big(\widehat{\btheta}_{a,\ell},\widehat{\btheta}_{b,\ell}|\bD_\ell\big)\big\} - \tfrac{1}{2}\big(\btheta_a-\widehat{\btheta}_{a,\ell}\big)^t \widehat{\bA}_{a,\ell}\big(\btheta_a-\widehat{\btheta}_{a,\ell}\big) \nonumber\\[7pt]
&&  -\;\tfrac{1}{2}\big(\btheta_{b,\ell}-\widehat{\btheta}_{b,\ell}\big)^t \widehat{\bA}_{b,\ell}\big(\btheta_{b,\ell}-\widehat{\btheta}_{b,\ell}\big)
-\big(\btheta_a-\widehat{\btheta}_{a,\ell}\big)^t \widehat{\bA}_{ab,\ell}\big(\btheta_{b,\ell}-\widehat{\btheta}_{b,\ell}\big) \nonumber
\end{eqnarray}
where we left out the third and higher order approximations, and where $\widetilde{\bA}_{b,\ell}$ and $\widetilde{\bA}_{ab,\ell}$ denote the matrices of minus the second derivatives of $\log \{p(\btheta|\bD)\}$ with respect to the components of $\btheta_{b,\ell}$ and with respect to both $\btheta_{a}$ and $\btheta_{b,\ell}$, respectively. 
As before we assume the usual zero-mean Gaussian priors for $\btheta_a$, and zero-mean Gaussian priors for $\btheta_{b,\ell}$ with inverse covariance matrices $\bLambda_{b,\ell}$ in $p_\ell(\btheta_{b,\ell})$ and $\bLambda_b$ in $p(\btheta_{b,\ell})$ for every $\ell$, so independent of $\ell$ (this assumption is not required, but makes notation simpler). Insertion of the above approximations for the log posterior densities and the prior densities of the parameters into (\ref{eq:linkyonly}),  gives
\begin{eqnarray*}
&& \big(\btheta_a-\widehat{\btheta}_a\big)^t \widehat{\bA}_a\big(\btheta_a-\widehat{\btheta}_a\big) 
+
 \sum_{\ell=1}^L \big(\btheta_{b,\ell}-\widetilde{\btheta}_{b,\ell}\big)^t \widetilde{\bA}_{b,\ell}\big(\btheta_{b,\ell}-\widetilde{\btheta}_{b,\ell}\big) 
  + \;2\big(\btheta_a-\widehat{\btheta}_a\big)^t\sum_{\ell=1}^L \widetilde{\bA}_{ab,\ell}\big(\btheta_{b,\ell}-\widetilde{\btheta}_{b,\ell}\big)
\nonumber\\
&& \hspace{5mm}\approx\;
 \sum_{\ell=1}^L 
 \big(\btheta_a-\widehat{\btheta}_{a,\ell}\big)^t \widehat{\bA}_{a,\ell}\big(\btheta_a-\widehat{\btheta}_{a,\ell}\big) 
 +\btheta_a^t\Big(\bLambda_a-\sum_{\ell=1}^L\bLambda_{a,\ell}\Big)\btheta_a \;+\;\sum_{\ell=1}^L 
  \big(\btheta_{b,\ell}-\widehat{\btheta}_{b,\ell}\big)^t \widehat{\bA}_{b,\ell}\big(\btheta_{b,\ell}-\widehat{\btheta}_{b,\ell}\big)
\\
&& \hspace{10mm} 
   \; + \; \sum_{\ell=1}^L \btheta_{b,\ell}^t\big(\bLambda_b-\bLambda_{b,\ell}\big)\btheta_{b,\ell}  \;+\; 2 \sum_{\ell=1}^L 
 \big(\btheta_a-\widehat{\btheta}_{a,\ell}\big)^t \widehat{\bA}_{ab,\ell}\big(\btheta_{b,\ell}-\widehat{\btheta}_{b,\ell}\big)
 +B \nonumber
\end{eqnarray*}
with $B$ representing a term that does not depend on $\btheta$. Both sides in the above  equation are quadratic functions of $\btheta$, and this equation must hold for all $\btheta$ in a neighborhood of $\widehat{\btheta}$. We must therefore equate the coefficients on either side of the specific linear and quadratic terms. Identification of the quadratic terms in $\btheta$ leads to the following equations: 
\begin{eqnarray}
\widehat{\bA}_a\approx\sum_{\ell=1}^L \widehat{\bA}_{a,\ell}+\bLambda_a-\sum_{\ell=1}^L\bLambda_{a,\ell},
~~~~~~~
\widetilde{\bA}_{b,\ell}\approx\widehat{\bA}_{b,\ell}+\bLambda_b-\bLambda_{b,\ell},
~~~~~~~
\widetilde{\bA}_{ab,\ell}\approx\widehat{\bA}_{ab,\ell},
\label{eq:with_nuisance_1}
\end{eqnarray}
whereas identification of the linear terms in $\btheta$ gives the two equations
\begin{eqnarray}
 \widehat{\bA}_a\widehat{\btheta}_a
 +\sum_{\ell=1}^L \widetilde{\bA}_{ab,\ell}\widetilde{\btheta}_{b,\ell}
&\approx&
 \sum_{\ell=1}^L 
 \widehat{\bA}_{a,\ell}\widehat{\btheta}_{a,\ell} 
+\sum_{\ell=1}^L 
  \widehat{\bA}_{ab,\ell}\widehat{\btheta}_{b,\ell},
\nonumber \\[7pt]
\widetilde{\bA}_{b,\ell}\widetilde{\btheta}_{b,\ell} 
+(\widetilde{\bA}_{ab,\ell})^t\widehat{\btheta}_a
&\approx&
\widehat{\bA}_{b,\ell}\widehat{\btheta}_{b,\ell}
 +(\widehat{\bA}_{ab,\ell})^t\widehat{\btheta}_{a,\ell}. \nonumber
\end{eqnarray}
These equations can again be solved, giving formulae for the estimators $\widehat{\btheta}_a$ and $\widetilde{\btheta}_{b,\ell}$ that would be obtained based on the combined data in terms of estimators in the separate centers. For the core parameters one finds:  
\begin{eqnarray}
&&\hspace*{-5mm}
\widehat{\btheta}_a 
\approx\Big( \widehat{\bA}_a
 -\sum_{\ell=1}^L \widehat{\bA}_{ab,\ell}(\widetilde{\bA}_{b,\ell})^{-1}(\widetilde{\bA}_{ab,\ell})^t
 \Big)^{\!-1}  \sum_{\ell=1}^L \Big\{
  \Big(
 \widehat{\bA}_{a,\ell}
 -
 \widehat{\bA}_{ab,\ell}(\widetilde{\bA}_{b,\ell})^{-1}(\widehat{\bA}_{ab,\ell})^t\Big)\widehat{\btheta}_{a,\ell} 
  + 
  \widehat{\bA}_{ab,\ell}\Big(\one
  -
(\widetilde{\bA}_{b,\ell})^{-1}\widehat{\bA}_{b,\ell}\Big)\widehat{\btheta}_{b,\ell}
\Big\}~~  \nonumber
\end{eqnarray}
with $\one$ the $d_2\times d_2$ unit matrix and
the expressions for the matrices $\widehat{\bA}_a$ and $\widetilde{\bA}_{b,\ell}$ as given in (\ref{eq:with_nuisance_1}). If required, the full set estimators for the nuisance parameters can subsequently be calculated from 
\begin{eqnarray}
\widetilde{\btheta}_{b,\ell} 
\approx\big(\widetilde{\bA}_{b,\ell}\big)^{-1}\Big\{\widehat{\bA}_{b,\ell}\widehat{\btheta}_{b,\ell}
 + \big(\widehat{\bA}_{ab,\ell}\big)^t\big(\widehat{\btheta}_{a,\ell} 
 -
 \widehat{\btheta}_a\big)\Big\}. \nonumber
 \end{eqnarray}
We conclude that also if nuisance parameters are different across centers, there is no need to combine the data subsets; we can approximate all relevant full set estimates {\em a posteriori} from the estimates obtained from the subsets. 

The above equations are seen to simplify if we choose identical nuisance parameter priors throughout, i.e., $\bLambda_{b,\ell}\approx\bLambda_b$  for all $\ell$. In that case $\widetilde{\bA}_{b,\ell}=\widehat{\bA}_{b,\ell}$, and hence 
\begin{eqnarray}
&&\hspace*{-5mm}
\widehat{\btheta}_a 
\approx\Big( \widehat{\bA}_a
 -\sum_{\ell=1}^L \widehat{\bA}_{ab,\ell}(\widehat{\bA}_{b,\ell})^{-1}(\widehat{\bA}_{ab,\ell})^t
 \Big)^{\!-1}\!\sum_{\ell=1}^L
  \Big(
 \widehat{\bA}_{a,\ell}
 -
 \widehat{\bA}_{ab,\ell}(\widehat{\bA}_{b,\ell})^{-1}(\widehat{\bA}_{ab,\ell})^t\Big)\widehat{\btheta}_{a,\ell}.  \nonumber
\end{eqnarray}

In Section \ref{sec:simstudies} an example is given in which the data in three centers follow a logistic regression model in which the center specific intercepts are the nuisance parameters and the remaining regression parameters are the parameters of interest. This results in an aggregated model that contains an intercept for each center population.

\section{Bayesian regression and subset inference}
\label{Sec:GLM}

\subsection{Bayesian regression}

The most relevant problem areas in medicine are those where each data point is a pair $(\bX_{\ell i},Y_{\ell i})$ of  input vectors $\bX_{\ell i}$ (the covariates) and output values $Y_{\ell i}$ (e.g., clinical outcome or treatment response). A parametrized model relates $\bX_{\ell i}$ to $Y_{\ell i}$, and the available data are used to infer the parameters for this model. In this section the BFI methodology described in Section \ref{sec:BFI} is applied to parametric regression models, and associated challenges such as heterogeneity across the populations are addressed.   
Suppose that $Y_{\ell i}|(\bX_{\ell i}=\bx_{\ell i},\btheta_1)$ and $\bX_{\ell i}|\btheta_2$ have densities $y|\bx,\btheta_1 \to p(y|\bx,\btheta_1)$ and $\bx|\btheta_2 \to p(\bx|\btheta_2)$, respectively, so that for $\btheta^t=(\btheta_1^t,\btheta_2^t)$ with unknown parameter vectors $\btheta_1\in\mathbb{R}^{ d_1}$ and $\btheta_2\in\mathbb{R}^{ d_2}$  we seek to infer:
$y,\bx|\btheta \to p(y,\bx|\btheta) \;=\; p(y|\bx,\btheta_1) p(\bx|\btheta_2).$
An example is the class of generalized linear models (GLMs). In GLMs the outcome variable $Y_{\ell i}$ is related to the covariates $\bX_{\ell i}$ only via the linear predictor $\bbeta^t \bX_{\ell i}$. By setting the first covariate equal to 1, an intercept can be included trivially in the model. Specially, for $h$ the link function, 
\begin{align*}
h(\mathbb{E}(Y_{\ell i}|\bX_{\ell i},\etab,\bbeta)) = \bbeta^t \bX_{\ell i},
\end{align*}
where $\bbeta$ is the vector of unknown regression parameters and $\etab$ a vector of unknown nuisance parameters (e.g., the variance of the outcome variable in a linear regression model). For this GLM the parameter $\btheta_1$ is defined as $\btheta_1=(\etab^t,\bbeta^t)^t$.

Let $\bD_\ell$ be the $\ell^{th}$ data subset, and $\bD$ be the union of the $L$ subsets:
\begin{eqnarray}
\bD_\ell=\{(\bx_{\ell 1},y_{\ell 1}),\ldots,(\bx_{\ell n_\ell},y_{\ell n_\ell})\},~~~~~~~~n_\ell=|\bD_\ell|,~~~~~~~~\bD=\bigcup_{\ell=1}^L \bD_\ell. \nonumber
\end{eqnarray}
As in the previous section, we aim to express the outcome of inference on the parameters in the combined set $\bD$ in terms of the outcomes of $L$ separate inferences on the constituent sets $\bD_\ell, \ell=1,\ldots,L$.  

For mathematical simplicity in subsequent expressions we assume statistically independent $\btheta_1$ and $\btheta_2$, i.e., $p(\btheta_1,\btheta_2)=p(\btheta_1)p(\btheta_2)$ and $p_\ell(\btheta_1,\btheta_2)=p_\ell(\btheta_1)p_\ell(\btheta_2)$ for all $\ell$. For the combined data set $\bD$ and for the subsets $\bD_\ell$  the log posterior distribution is then given by, respectively, 
\begin{eqnarray}
\log \big\{p(\btheta|\bD)\big\} &=& \log \big\{p(\btheta)\big\} + \sum_{\ell=1}^L\sum_{i=1}^{n_\ell} \log \big\{p(y_{\ell i},\bx_{\ell i}|\btheta)\big\} -\log \big\{Z(\bD)\big\}  
\\
&=&  \log \big\{p(\btheta_1)\big\} + \sum_{\ell=1}^L\sum_{i=1}^{n_\ell} \log \big\{p(y_{\ell i}|\bx_{\ell i},\btheta_1)\big\} + \log \big\{p(\btheta_2)\big\} + \sum_{\ell=1}^L\sum_{i=1}^{n_\ell} \log \big\{p(\bx_{\ell i}|\btheta_2)\big\} -\log \big\{Z(\bD)\big\} \nonumber
\label{eq:logp_D} \\
\log \big\{p_\ell(\btheta|\bD_\ell)\big\}  &=&  \log \big\{p_\ell(\btheta)\big\} + \sum_{i=1}^{n_\ell} \log \big\{p(y_{\ell i},\bx_{\ell i}|\btheta)\big\} -\log \big\{Z_\ell(\bD_\ell)\big\} 
\\
&=&  \log \big\{p_\ell(\btheta_1)\big\} +\sum_{i=1}^{n_\ell} \log \big\{p(y_{\ell i}|\bx_{\ell i},\btheta_1)\big\} + \log \big\{p_\ell(\btheta_2)\big\} + \sum_{i=1}^{n_\ell} \log \big\{p(\bx_{\ell i}|\btheta_2)\big\} -\log \big\{Z_\ell(\bD_\ell)\big\}. \nonumber
\label{eq:logpD_l}
\end{eqnarray}
So, the log posterior densities are decomposed into terms that depend on either $\btheta_1$, or on $\btheta_2$ (or neither), and  maximization with respect to $\btheta_1$ and $\btheta_2$ to obtain their MAP estimators can be performed independently. Similarly as in (\ref{eq:link}), the link between the log posterior density for the combined data set and for the subsets takes the form 
\begin{eqnarray}
\label{eq:full_into_subsets}
\log \big\{p(\btheta|\bD)\big\} &=& \sum_{\ell=1}^L \log \big\{p_\ell(\btheta|\bD_\ell)\big\}+\log\Bigg(\frac{p(\btheta_1)}{\prod_{\ell=1}^L p_\ell(\btheta_1)}\Bigg)+\log\Bigg(\frac{p(\btheta_2)}{\prod_{\ell=1}^L p_\ell(\btheta_2)}\Bigg)-\log\Bigg(\frac{Z(\bD)}{\prod_{\ell=1}^L Z_\ell(\bD_\ell)}\Bigg). 
\end{eqnarray}

Again  the log posterior densities $\log \{p(\btheta|\bD)\}$ and $\log \{p_\ell(\btheta|\bD_\ell)\}$ can for sufficiently large $n$ be approximated by quadratic expansions for $\btheta$ near the MAP estimators $\widehat{\btheta}$ and $\widehat{\btheta}_\ell$, leading to Gaussian approximations for the posterior densities themselves. Since the log posterior densities can both be decomposed in terms dependent on either $\btheta_1$ or $\btheta_2$ (there are no mixture terms), the matrices $\widehat{\bA}$ and $\widehat{\bA}_\ell$ are diagonal block matrices of the form 
\begin{eqnarray}
\widehat{\bA}=\left(\!\begin{array}{cc}
\widehat{\bA}_1 & {\bf 0}\\
{\bf 0} & \widehat{\bA}_2\end{array}\!
\right),~~~~~~~~\widehat{\bA}_\ell=\left(\!\begin{array}{cc}
\widehat{\bA}_{1,\ell} & {\bf 0}\\
{\bf 0} & \widehat{\bA}_{2,\ell}\end{array}\!
\right), \nonumber
\end{eqnarray}
in which the blocks $\{\widehat{\bA}_1,\widehat{\bA}_{1,\ell}\}$ and $\{\widehat{\bA}_2,\widehat{\bA}_{2,\ell}\}$ represent the minus  curvature matrices for $\btheta_1$ and  $\btheta_2$, respectively. 
The quadratic log-likelihood approximation for $p(\btheta|\bD)$ is
\begin{eqnarray}
\log \big\{p\big(\btheta|\bD\big)\big\} \;\approx\; \log \big\{p\big(\widehat{\btheta}|\bD\big)\big\} -\tfrac{1}{2}(\btheta_1-\widehat{\btheta}_1)^t \widehat{\bA}_1\big(\btheta_1-\widehat{\btheta}_1\big)  -\tfrac{1}{2}\big(\btheta_2-\widehat{\btheta}_2\big)^t \widehat{\bA}_2(\btheta_2-\widehat{\btheta}_2), \nonumber
\end{eqnarray}
and a similar expression holds for $p_\ell(\btheta|\bD_\ell)$. If all prior parameter densities are chosen to be Gaussian with mean zero and inverse covariance matrices $\bLambda_1$ and $\bLambda_{1,\ell}$ for $\btheta_1$ and $\bLambda_2$ and $\bLambda_{2,\ell}$ for $\btheta_2$, then insertion of these quadratic approximations %(\ref{eq:logp}, \ref{eq:logpl}) 
into (\ref{eq:full_into_subsets}) 
gives, for $k=1,2$
\begin{eqnarray}
\big(\btheta_k-\widehat{\btheta}_k\big)^t \widehat{\bA}_k\big(\btheta_k-\widehat{\btheta}_k\big) \;\approx\; 
\btheta_k^t \Big(\bLambda_k -\sum_{\ell=1}^L \bLambda_{k,\ell}\Big) \btheta_k+
\sum_{\ell=1}^L 
\big(\btheta_k-\widehat{\btheta}_{k,\ell}\big)^t \widehat{\bA}_{k,\ell}\big(\btheta_k-\widehat{\btheta}_{k,\ell}\big) +B_k,
\nonumber
\end{eqnarray}
with $B_1$ and $B_2$ not dependent on $\btheta$. Expressions for $\widehat{\btheta}_1, \widehat{\btheta}_2, \widehat{\bA}_1$ and $\widehat{\bA}_2$ in terms of their subset analogous can now be extracted similarly as in  (\ref{eq:recombination_formulae}), giving
\begin{eqnarray}
\hspace*{-5mm} && \widehat{\bA}_1 \approx \sum_{\ell=1}^L \widehat{\bA}_{1,\ell}+\bLambda_1-\sum_{\ell=1}^L \bLambda_{1,\ell},~~~~~~~~~
\widehat{\btheta}_1\approx  \big(\widehat{\bA}_1\big)^{-1}\sum_{\ell=1}^L \widehat{\bA}_{1,\ell}\widehat{\btheta}_{1,\ell},
\label{eq:recover_theta1}
\\
\hspace*{-5mm} && \widehat{\bA}_2 \approx \sum_{\ell=1}^L \widehat{\bA}_{2,\ell}+\bLambda_2-\sum_{\ell=1}^L \bLambda_{2,\ell},~~~~~~~~~
\widehat{\btheta}_2\approx \big(\widehat{\bA}_2\big)^{-1}\sum_{\ell=1}^L \widehat{\bA}_{2,\ell}\widehat{\btheta}_{2,\ell}.
\label{eq:recover_theta2}
\end{eqnarray}
From these formulae one can see  that the distribution of the covariates does not affect the estimator $(\widehat{\btheta}_1, \widehat{\bA}_1)$; i.e., it does not affect the estimator of the regression parameters. If independence between $\btheta_1$ and $\btheta_2$ can not be assumed, the matrices $\widehat{\bA}$ and $\widehat{\bA}_\ell$ are not necessarily block matrices and the expressions in (\ref{eq:recover_theta1}) and (\ref{eq:recover_theta2}) do not hold anymore. The general theory in Section \ref{sec:BFI} is still applicable and expressions for $\widehat{\btheta}$ and $\widehat{\bA}$ can be obtained from  (\ref{eq:recombination_formulae}).   

It is well known that there is a direct link between Bayesian inference for regression models and penalized regression. If one chooses a Gaussian prior with a diagonal inverse covariance matrix $\bLambda$ with all diagonal elements equal to $\lambda$, the MAP estimator for $\btheta$ is identical to the ridge penalized ML estimator for $\btheta$ with regularizer $\lambda$.\cite{Wu}  %(Wu et al.\ 2018\nocite{Wu}). 
The higher the value of $\lambda$, the stronger the penalization, or the smaller the variance of the prior distribution. 
When the sample size in one or more local subsets is small compared to the number of covariates in the model, the model in that particular center may be overfitted.  Increasing the value of the regularizer $\lambda_\ell$ (i.e., of the inverse variance of the prior in subset $\ell$) can prevent overfitting in the local models and increase the inference accuracy.  This is illustrated in Section \ref{sec:simstudies}. Note that in most ridge regression analyses only the regression (or association) parameters are penalized, not the intercept and the nuisance parameters. However, by assuming flat priors for the intercept and the nuisance parameters the two approaches are similar.

\subsection{Heterogeneity across populations}

So far we have assumed that the distributions of the covariates and the conditional distributions of the outcome given the covariates are the same in all centers. In practice this means that we assumed that the subsets were sampled from statistically identical populations. This assumption might be violated, for instance if the different centers represent different nationalities or hospitals. In that case, the $L$ subsets are sampled from structurally different populations, and the distributions of the covariates and/or of the (conditional) outcomes may well differ across centers. If we assume that the regression parameters $\btheta_1$ do not vary across centers, but that the parameters of the covariate distributions, $\btheta_2$, may do so, and choose simple priors of the form $p(\btheta)=p(\btheta_1)\prod_{\ell=1}^L p(\btheta_{2,\ell})$, then the calculations as before yield equation (\ref{eq:recover_theta1}) derived earlier for all estimators related to $\btheta_1$.  Estimation of $\btheta_1$ is hence not affected by having subset-specific parameters $\btheta_2$. Equation (\ref{eq:recover_theta2}), in contrast, is now replaced by 
\begin{eqnarray}
&&
\widetilde{\bA}_{2,\ell}\approx\widehat{\bA}_{2,\ell}+\bLambda_2-\bLambda_{2,\ell},~~~~~~~~~
\widetilde{\btheta}_{2,\ell}\approx\big(\widehat{\bA}_{2,\ell}+\bLambda_2-\bLambda_{2,\ell}\big)^{-1}
\widehat{\bA}_{2,\ell}\widehat{\btheta}_{2,\ell},
\nonumber
\end{eqnarray}
where $\widetilde{\btheta}_{2,\ell}$ is the MAP estimator of $\btheta_{2,\ell}$ found via inference on the full set $\bD$, and $\widetilde{\bA}_{2,\ell}$ denotes the matrix of minus second derivatives of $\log \{p(\btheta|\bD)\}$ with respect to $\btheta_{2,\ell}$. 

By comparing the local estimators $\widehat\btheta_{2,\ell}$ and considering the credible intervals as described in Subsection 2.3, the assumption of identical local populations may be verified.

\subsection{Sets of covariates differ across centers}
The actual set of available covariates may differ across centers, for instance because measuring and registering some medical patient characteristics is or was not part of the standard care in one or more centers. As a consequence, some local data sets may not contain data of some covariates.  Reconstructing the true (missing) observations is not always possible, because measurements may be time dependent (e.g., blood pressure), the patient may have died, and privacy legislation makes it sometimes impossible and certainly time-consuming. 

In general, if an individual covariate value is missing,  imputation methods can be applied. Common methods are single and multiple regression imputation\cite{Buuren}: %(van Buuren, 2018\nocite{Buuren}): 
a value for a missing covariate is predicted by a regression model that was fitted based on the observed covariates and outcome values. If in one of the centers or local data set, a covariate is completely missing a similar strategy can be applied. A prediction model for the unobserved covariate is fitted in the centers in which the covariate is measured, the models are combined by means of the BFI strategy in order to obtain one regression model (according to the formulae above) and subsequently used to predict and impute the covariate values in the centers where the covariate was not measured. After imputation, the final BFI model can be constructed as before. This strategy only works if the simultaneous distributions of the covariates are identical across centers.

\section{Simulation Studies and Application}
\label{sec:simstudies}

\subsection{Study aims}

In this section we describe the results of several simulation studies aimed at quantifying the performance of the BFI methodology. More specifically, we study the agreement between the inference results of the BFI strategy and those obtained based on the combined data of the $L$ subsets. 

Regression models are employed for multiple purposes, e.g., for predicting outcomes of new patients or subjects, or for studying association between a factor and the outcome. In association studies, the estimated regression parameters are of main interest, whereas in prediction studies the focus is on outcome prediction. If the estimates of the regression parameters obtained with the BFI methodology and those based on the combined data are close, predictions based on the two models will be very similar as well. However, if the covariates are correlated, similar predictions may also be obtained if the regression parameters differ. Therefore, we are interested in prediction agreement as well.  

We use an existing real life data set for illustrating the BFI methodology. The data sets come from three locations and have different sample sizes. The outcome of interest is binary and, thus, a logistic regression model is fitted. Data of four covariates are available to predict the outcome. In these three data-sets the number of events per covariate (EPV) is below 10 or approximately equal to 10. To overcome overfitting a minimum EPV of 10 is often advised.

We have access to the data in all three locations, and from that perspective this example can be seen as a toy-example to illustrate the methodology and is not meant for deriving any practical conclusions. Based on the data, we aim to study the performance of the BFI method in the following settings:
\begin{enumerate}
\item  Homogeneity across populations (local centers). The parameters across the three locations are equal.
\item Low sample sizes in local centers. The sample sizes in the subsets are small compared to the number of covariates in the model. 
\item Heterogeneity across local centers. The intercepts of the models in the different locations differ.  
\end{enumerate}   
More information on the data is given in the next subsection and in Appendix A.

\subsection{Real life data}

Our data set consists of data of trauma patients from different hospitals. The outcome variable of interest is mortality (binary) and the covariates are age, sex, the Injury Severity Score (ISS, 
ranging from 1 (low) to 75 (high)), the Glasgow Coma Scale (GCS, which expresses the level of consciousness, ranging from 3 (low) to 15 (high)). We have data from three (types of) medical centers: a peripheral hospital without a neuro-surgical unit (NSU),  a peripheral hospital with NSU, and an academic medical center.\cite{Draaisma} The sample sizes in the three centers are 49, 106 and 216 patients, respectively. Because each subset comes from a specific hospital type, the patient populations and characteristics of the patients' traumas may be different, i.e., the distributions of the covariates and outcome variable may vary across the centers. This is certainly the case for the mortality rate: 43\% in the peripheral hospital without NSU, 40\% in the peripheral hospital with NSU, and 22\% in the academic medical center. Also the median ISS rates differ, from 29 in the academic hospital to 41 in the peripheral hospital without NSU. The median age, percentage females and median GCS are similar across the centers (they equal 29 years, 27\% females and 12 for GCS in the combined data set). Even if populations differ, the regression parameters of the covariates may equal. 
More details on the data are given in  Appendix A. 

We assume that the data in the combined and in the local data sets come from Bayesian logistic regression models (with possibly different parameters). Let $Y_{\ell i}$ be a stochastic variable that indicates whether patient $i$ from hospital $\ell$ died ($Y_{\ell i}=1$) or not ($Y_{\ell i}=0$), and let 
$\bX_{\ell i}$ be a vector with the covariates for this patient: 1 (for the intercept), age, sex (0: males, 1: females), ISS, and GCS. We assume that, for a patient, the probability that $Y_{\ell i}=1$, given his covariates and given the vector $\bbeta$ is given by
\begin{align*}
\rP(Y_{\ell i}=1|\bX_{\ell i},\bbeta)  = \frac{\exp(\bbeta^t \bX_{\ell i})}{1+\exp(\bbeta^t \bX_{\ell i})},   
\end{align*}
with the prior distribution $\bbeta \sim N_5(\bf{0}, \bLambda^{-1})$,
a 5 dimensional Gaussian distribution with mean zero and a diagonal covariance matrix.

We combined the data, normalized every covariates (subtracted the sample mean and divided by the sample standard deviation of the full data set), and fitted a Bayesian logistic regression model in the combined data set.  
The diagonal elements of the covariance matrix are set equal to $10$ (the diagonal elements of $\bLambda=\bLambda_\ell$ equal $\lambda=0.1$). The MAP-estimated regression parameters with their estimated standard deviations for the combined set and in the three different subsets are given in Table \ref{tab:local}. One should always be careful trying to interpret the estimates of the regression parameters in a multivariable model, as the covariates may be correlated. However, from the estimates in the centers as well as those obtained after combining the data, we can carefully conclude that the mortality risk increases with age and ISS, but decreases with increasing GCS, which seem reasonable. The regression coefficient for sex is negative, but the corresponding standard deviation is relatively large.
%\begin{table}
%\begin{center}
%\begin{tabular}{|l|r|r|r|r|r|}
%\hline
%   & intercept & sex & age & ISS  & GCS \\
%\hline
% peripheral no NSU & $-2.82~~ (0.94)$ & $-1.74~~ (0.90)$ &  $1.54~~ (0.56)$ &  $1.90~~ (0.70)$ & $-2.06~~ (0.65)$ \\
% peripheral with NSU & $-0.97~~ (0.33)$ & $-0.34~~ (0.31)$ & $1.90~~ (0.45)$ &  $0.52~~ (0.36)$ & $-1.82~~ (0.40)$\\
% academic hospitals & $-2.22~~ (0.36)$ &  $0.07~~ (0.25)$ &  $1.27~~ (0.29)$ &  $0.30~~ (0.26)$ & $-2.36~~ (0.40)$\\
%\hline
% Combined ($\bbeta_c$) & $-1.70~~ (0.22)$ & $-0.15~~ (0.18)$ &  $1.36~~ (0.20)$ & $0.55~~ (0.19)$ & $-1.98~~ (0.24)$\\
%\hline
%\end{tabular}
%\caption{
%Estimates of $\bbeta$ (and their standard deviations) in each subset and obtained after first combining the data. The diagonal elements of the covariance matrix are equal to $10$. } 
%\label{tab:local}
%\end{center}
%\end{table}

\begin{table}
\begin{center}
\begin{tabular}{|l|r|r|r|r|r|}
\hline
   & intercept & sex & age & ISS  & GCS \\
\hline
 peripheral without NSU & $-2.82~~ (0.94)$ & $-1.74~~ (0.90)$ &  $1.54~~ (0.56)$ &  $1.90~~ (0.70)$ & $-2.06~~ (0.65)$ \\
 peripheral with NSU & $-0.97~~ (0.33)$ & $-0.34~~ (0.31)$ & $1.90~~ (0.45)$ &  $0.52~~ (0.36)$ & $-1.82~~ (0.40)$\\
 academic hospital & $-2.22~~ (0.36)$ &  $0.07~~ (0.25)$ &  $1.27~~ (0.29)$ &  $0.30~~ (0.26)$ & $-2.36~~ (0.40)$\\
 \hline
 Combined ($\bbeta_c$) & $-1.70~~ (0.22)$ & $-0.15~~ (0.18)$ &  $1.36~~ (0.20)$ & $0.55~~ (0.19)$ & $-1.98~~ (0.24)$\\
 BFI ($\bbeta_b$) & $-1.51~~ (0.22)$ & $-0.12~~ (0.18)$ & $1.23~~ (0.21)$ & $0.51~~ (0.19)$& $-1.74~~ (0.24)$\\
\hline
\end{tabular}
\caption{
Estimates of the regression parameters  in each center (including estimated standard deviations), after combining the data and based on the BFI strategy. Here $\lambda=0.1$. } 
\label{tab:local}
\end{center}
\end{table}

\subsection{Homogeneity across population}
\label{sub:homo} 

\noindent
{\bf{Simulation procedure}}\\
In this subsection we study the performance of the BFI methodology for the case where the data in the subsets come from populations with identical characteristics, i.e., with identical covariate distributions and identical true parameter values (including those of the regression parameters). Since this does not seem to hold in our data set (see Table \ref{Tab:data}), we first randomly assign each patient to one of the three subsets, keeping the sample sizes for the three subsets fixed. After this randomization we can indeed assume that the three patient groups have the same covariate distributions and regression parameters. 
We then perform regression analyses at subset level and at full data set level, and quantify  the agreement between the results obtained on the full combined data set versus those obtained via the BFI protocol for integrating the outcomes of the subset regressions. This procedure is repeated multiple  times. Below we describe the quantities used for measuring agreement. For the prediction comparisons some extra steps are taken to ensure that estimation and prediction are not based on the same data. In every cycle and every subset, 10\% of the patients are randomly left out of the estimation procedure. After obtaining the estimates for the regression coefficients with the BFI strategy, mortality probabilities are predicted for the 10\% patients that were left out (based on their covariates and estimated regression parameters). More details are given below. 

\bigskip

\noindent
{\bf{Quantities to measure agreement}}\\
The outcome variable is binary (mortality) and, therefore, a logistic regression model is fitted. The parameter of interest is the vector of regression parameters $\bbeta$ (including the intercept). For simplicity of interpretation of the regression parameters and the choice of the prior distribution, we first normalize every covariate by subtracting from each covariate value in the data set the sample mean of the covariate values in the combined data set and dividing by the sample standard deviation (computed in the combined data set). This is done for each covariate. Note that this sample mean and standard deviation in the combined data set can be computed without combining all local data, because they are functions of the first and second sample moments in every local subset only (and of the local sample size). Therefore, the local data themselves do not need to be combined and standardization is also possible in practice. 

Let $(\widehat{\bbeta}_b, \widehat{\bA}_b)$ and $(\widehat{\bbeta}_c, \widehat{\bA}_c)$ denote the estimates of $(\bbeta, \bA)$ based on the BFI strategy and found after combining all data (the subscript $b$ stands for BFI, $c$ for combined data). To quantify the agreement between the regression estimates, the mean squared error (MSE) is computed for every regression coefficient with the formula:  
\begin{align*}
MSE_{\beta_k} = \frac{1}{M}\sum_{m=1}^M  \Big(\widehat{\beta}_{b k}^{(m)}-\widehat{\beta}_{c k}\Big)^2. 
%\label{eq:MSEbeta}
\end{align*} 
Here $k=1,\ldots,d$ labels the $d$ components of the vector $\bbeta$, and the superscript $m$ labels the $M$ independent randomization cycles of the patients over the three subsets. Note that the values $\widehat{\beta}_{ck}$ are the same in every cycle (i.e., is independent of $m$), since this estimate is based on the combined data. 
A small value of $MSE_{\beta_k}$ means that there is hardly any loss when computing the MAP estimates with the BFI method (i.e., without combining all the data). Furthermore, the MSE is calculated for the square root of every diagonal element of minus the inverse of the curvature matrix (representing estimators of the expected errors in the regression coefficients):
\begin{align*}
MSE_{A_k} = \frac{1}{M}\sum_{m=1}^M  \Bigg\{\sqrt{\Big\{\big(\widehat{\bA}_{b}^{(m)}\big)^{-1}\Big\}_{kk}}
-\sqrt{\big(\widehat{\bA}_{c}^{-1}\big)_{kk}}~\Bigg\}^2. 
%\label{eq:MSEA}
\end{align*} 
For testing the accuracy of BFI's patient outcome predictions we use similar quantities. For the 10\% individuals that were left out from the estimation part, the prediction probabilities are estimated. In each cycle $m$, for an individual $i$ from center $\ell$, these probabilities are computed as  
\begin{eqnarray}
\widehat{p}_{b\ell i}^{(m)} = \frac{\exp\Big\{\big(\widehat{\bbeta}^{(m)}_{b}\big)^t \bx_{\ell i}\Big\}}{1+\exp\Big\{\big(\widehat{\bbeta}^{(m)}_{b}\big)^t \bx_{\ell i}\Big\}},~~~~~~~~~~~~ 
\widehat{p}_{c\ell i}^{(m)} = \frac{\exp\Big\{\big(\widehat{\bbeta}^{(m)}_{c}\big)^t \bx_{\ell i}\Big\}}{1+\exp\Big\{\big(\widehat{\bbeta}^{(m)}_{c}\big)^t \bx_{\ell i}\Big\}}.
\label{eq:probabilities}
\end{eqnarray}
The estimators $\widehat{\bbeta}^{(m)}_{b}$ and $\widehat{\bbeta}^{(m)}_{c}$ are both obtained from 90\% of the patients (a selection that is different for each cycle $m$, so both estimators depend on $m$). 
To compare the results from the two regression routes (regression after data integration  versus BFI recombination of subset results), we compute
\begin{align}
MSE_p = \frac{1}{37 M} \sum_{m=1}^M \sum_{\ell=1}^L \sum_{i\in S_\ell^{(m)}} \Big(\widehat p_{b\ell i}^{(m)}-\widehat p_{c\ell i}^{(m)}\Big)^2,
\label{exp:MSEp} 
\end{align}
where $S_\ell^{(m)}$ is the set of indices of individuals selected for prediction in subset $\ell$ in cycle $m$. The value of 37 is equal to 10\% of the total sample size, and hence gives the number of individuals selected for prediction in every cycle. 

\bigskip

\noindent
{\bf{Results: Parameter estimation}}\\
The prior distribution for the vector of regression parameters $\bbeta$ is chosen to be multivariate Gaussian with mean zero and a diagonal covariance matrices $\bLambda^{-1}=\bLambda_\ell^{-1}$ of which every element equals $\lambda$. In every subset $\bD_\ell$ we fit a logistic regression model with the same prior (i.e., $\bLambda_\ell=\bLambda$ for all $\ell$), and we aggregate the results by applying the formulae in (\ref{eq:recombination_formulae}). The $MSE_{\beta_k}$ are in the upper part of Table \ref{Table:MSEHomo}.

%\spacingset{1.7}
\begin{table}
\begin{center}
\begin{tabular}{|l|l|r|r|r|r|r|}
\hline
& & intercept  &  sex &  age & ISS   & GCS  \\
\hline \hline
$\lambda=0.1$ & $\widehat{\beta}_c ~~ (sd)$ & $-1.70~~ (0.22)$ & $-0.15~~ (0.18)$ & $1.36~~ (0.20)$ & $0.55~~ (0.19)$ & $-1.98~~ (0.24)$ \\
%& $\widehat{sd}(\widehat{\beta}_c)$ & 0.22  &   0.18 &    0.20  &   0.19 &  0.24 \\
& $MSE_{\beta_k}$ & $11.11$ & $0.60$ & $8.90$ & $0.96$ & $20.00$\\
& $MSE_{A_k}$ & $0.13$ & $0.25$ & $0.13$ & $0.11$ & $0.28$ \\
\hline
$\lambda=0.01$ & $\widehat{\beta}_c ~~ (sd)$ & $-1.72~~ (0.22)$ & $-0.15~~ (0.18)$ & $1.37~~ (0.21)$ & $0.55~~ (0.19)$ & $-2.00~~ (0.24)$ \\
%& $\widehat {sd}(\widehat\beta_c)$ & 0.22  &   0.18 &    0.21  &   0.19 &  0.24 \\
 & $MSE_{\beta_k}$ & $15.71$ & $0.99$ & $12.62$ & $1.70$ & $27.73$ \\
& $MSE_{A_k}$ & $0.43$ & $0.49$ & $0.30$ & $0.20$ & $0.21$ \\
\hline \hline
$\lambda_1=1, \lambda_2=0.1$ & $MSE_{\beta_k}$ & $16.60$  & $0.58$ & $13.05$ & $1.00$ & $27.56$\\
$\lambda_3=\lambda=0.01$ & $MSE_{A_k}$ & $1.55$ & $0.13$ & $1.12$ & $0.29$ & $2.85$ \\
\hline \hline
$\lambda=1$ & $\widehat{\beta}_c ~~ (sd)$ & $-1.58~~ (0.20)$ & $-0.13~~ (0.17)$ & $1.25~~ (0.19)$ & $0.54~~ (0.18)$ & $-1.84~~ (0.21)$\\
%& $\widehat {sd}(\widehat\beta_c)$ & 0.20  & 0.17 & 0.19 & 0.18 & 0.21 \\
& $MSE_{\beta_k}$ & $30.52$ & $0.42$ & $27.08$ & $1.90$ & $53.77$\\
& $MSE_{A_k}$ &  $15.91$ & $2.83$ & $10.97$ &  $4.02$ & $20.47$\\
\hline
$\lambda=0.1$ &  $MSE_{\beta_k}$ & $74.76$  & $1.90$ & $63.71$ & $5.31$ & $143.33$\\
& $MSE_{A_k}$ & $0.32$ & $2.15$ & $0.35$ & $0.37$ & $1.27$ \\
\hline
$\lambda=0.01$ &  $MSE_{\beta_k}$ & $183.89$ & $4.60$ & $143.92$ & $11.42$ & $308.51$ \\
& $MSE_{A_k}$ & $6.69$& $9.75$ & $3.92$ & $3.55$ & $2.03$ \\
\hline
\end{tabular}
\caption{
Results of the simulation study: estimates $\widehat \beta_c$, the $MSE_{\beta_k}$ (multiplied by $10^{3}$) and $MSE_{A_k}$ (multiplied by $10^{4}$). 1) for homogeneous populations (rows 1 and 2) as described in Subsection \ref{sub:homo}. The priors are zero-mean multivariate Gaussian distributions with the same covariance matrix in every subset and in the combined data set;\; 2) the priors in the subsets and the combined data set are taken equal to zero-mean multivariate Gaussian priors with different diagonal covariance matrices (row 3), with $\lambda_i$ the value in subset $i$ and $\lambda$ the value in the combined data set (Subsection \ref{sub:homo});\; 3) the situation in which the subsets are small (rows 4-6) as described  in Subsection \ref{sub:small}. In all cases $M=1000$. 
}
\label{Table:MSEHomo}
\end{center}
\end{table}

%\spacingset{2}
In this table we see that the estimates based on the BFI strategy are accurate. As expected, for higher values of $\lambda$ the estimates of the regression coefficients are closer to zero, and the estimated standard deviations are smaller. Furthermore, the $MSE$ values are also closer to zero for higher values of $\lambda$. However, better agreement between the estimates based on the BFI strategy and those found after combining the data, does not necessarily mean that the estimates are closer to the true values. 
 
\bigskip

\begin{figure}
\begin{center}
\includegraphics[width=5.9cm,height=7cm]{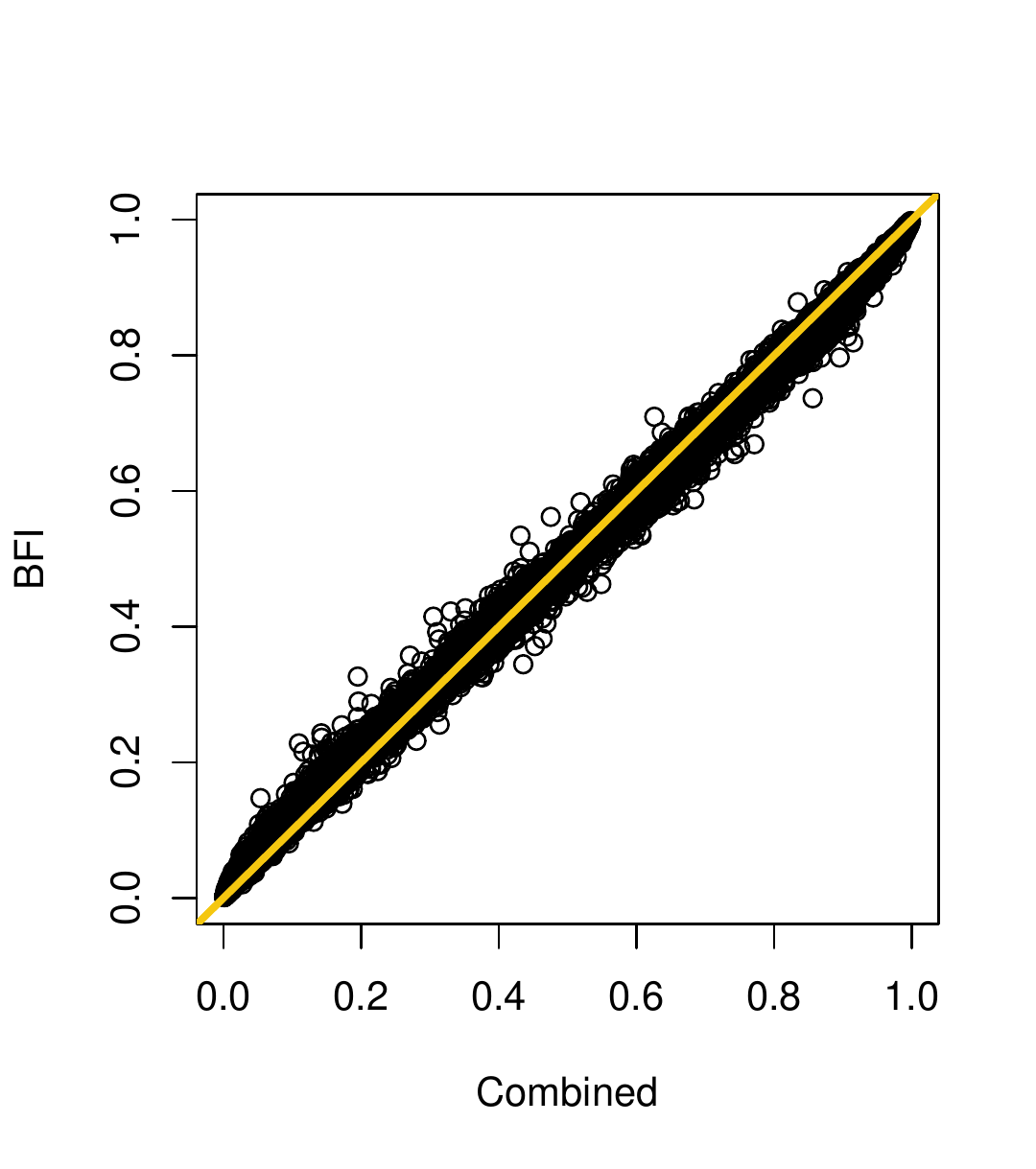}\;
\includegraphics[width=5.9cm,height=7cm]{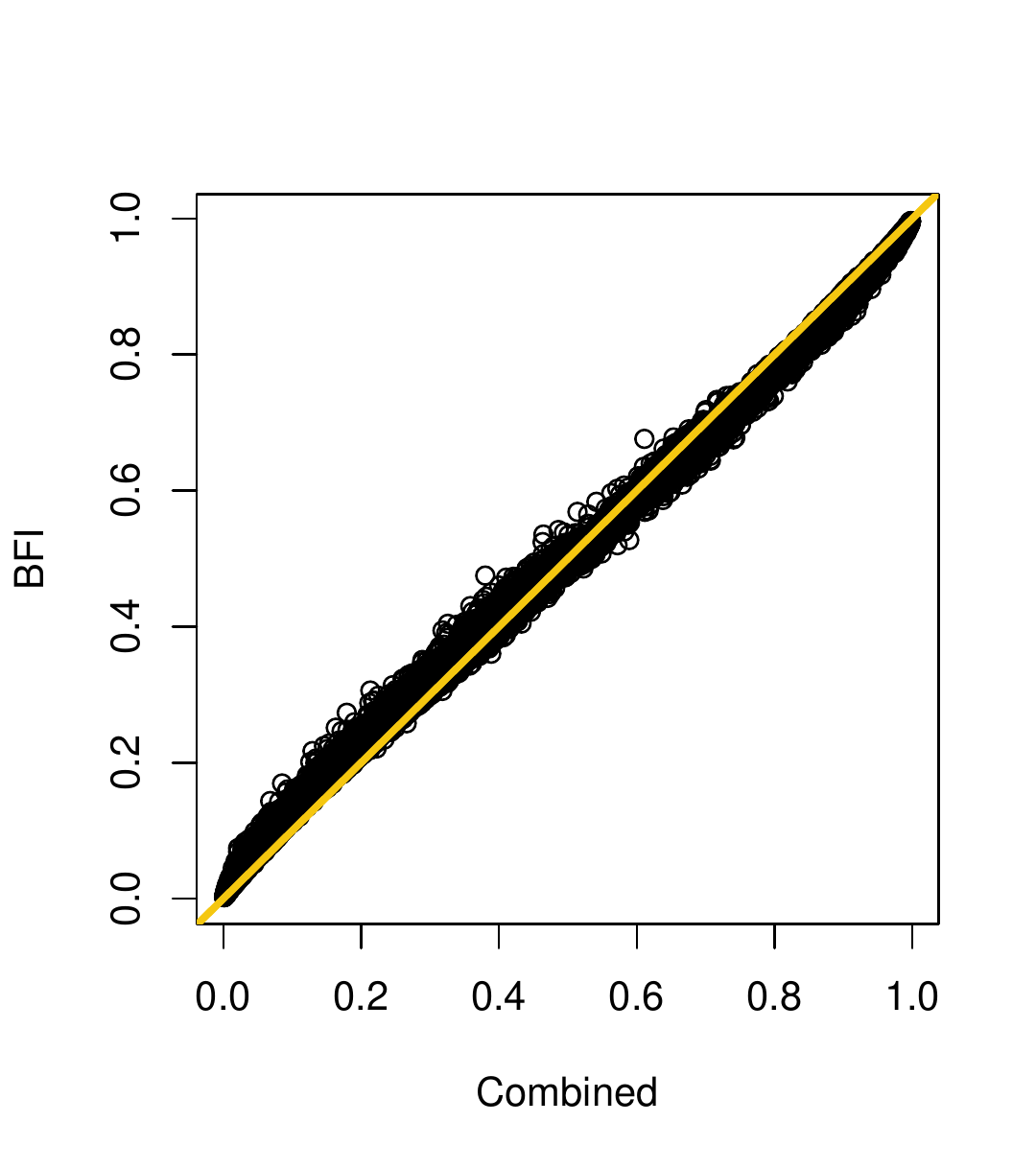}\;
\caption{Prediction probabilities based on the estimates $\widehat{\bbeta}_{b}$ (BFI, vertical axis) versus those based on $\widehat{\bbeta}_c$ (regression on combined data set, horizontal axis), for $\lambda=0.1$. Perfect agreement corresponds to all points being located on the diagonal. Left:  Homogeneous populations.  Right: Heterogenous populations. %Right: Missing covariate. 
In both plots the correlation $R^2=1.0$. 
}
\label{fig:PredProb}
\end{center}
\end{figure}

\noindent
{\bf{Results: Prediction}}\\
The BFI-estimated logistic regression model is used to predict mortality for new patients. To quantify its prediction performance we carried out simulation studies as described above: in each of the $M=1000$ simulation cycles 10\% of the patients were left out from estimation, and used instead to predict the mortality probability. The quantity $MSE_p$ in (\ref{exp:MSEp}) was calculated, resulting in the values $MSE_p=0.25\times 10^{-3}$ for $\lambda=0.01$, and $MSE_p=0.17 \times 10^{-3}$ for $\lambda=0.1$.  
In addition, in Figure \ref{fig:PredProb} (left) we show a scatter plot with the predicted probabilities based on the BFI model (vertical axis) and the combined data model (horizontal axis), for $\lambda=0.1$.
This plot shows excellent agreement. As expected for the present model definitions, the data points in Figure \ref{fig:PredProb} (left) are not distributed homogeneously along the diagonal, since Gaussian distributed uncertainty in parameter estimates is mapped nonlinearly to probabilities via the expressions in (\ref{eq:probabilities}). This effect is more clearly present when the local data sets are smaller (next subsection), where the estimate uncertainties are larger. 

\bigskip

\noindent
{\bf{Results: Parameter estimation with different priors}}\\
We also considered the situation in which the priors for the vector of regression parameters $\bbeta$ in the subsets are  different. We took multivariate Gaussian distributions with mean zero again, but with inverse covariance matrices with on the diagonal $\lambda_1=1, \lambda_2=0.1$ and $\lambda_3=0.01$ for the three subsets  and $\lambda=0.01$ for the combined data set (so the lower the sample size the stronger the regularisation). The $MSE_{\beta_k}$ are in row 3 of Table \ref{Table:MSEHomo}. The estimates based on the BFI strategy are accurate. We also performed a simulation study for mortality prediction. The results are very similar to those in Figure \ref{fig:PredProb} (left). 

\subsection{Small subsets} 
\label{sub:small}
\begin{figure}
\begin{center}
\includegraphics[width=5.9cm,height=7cm]{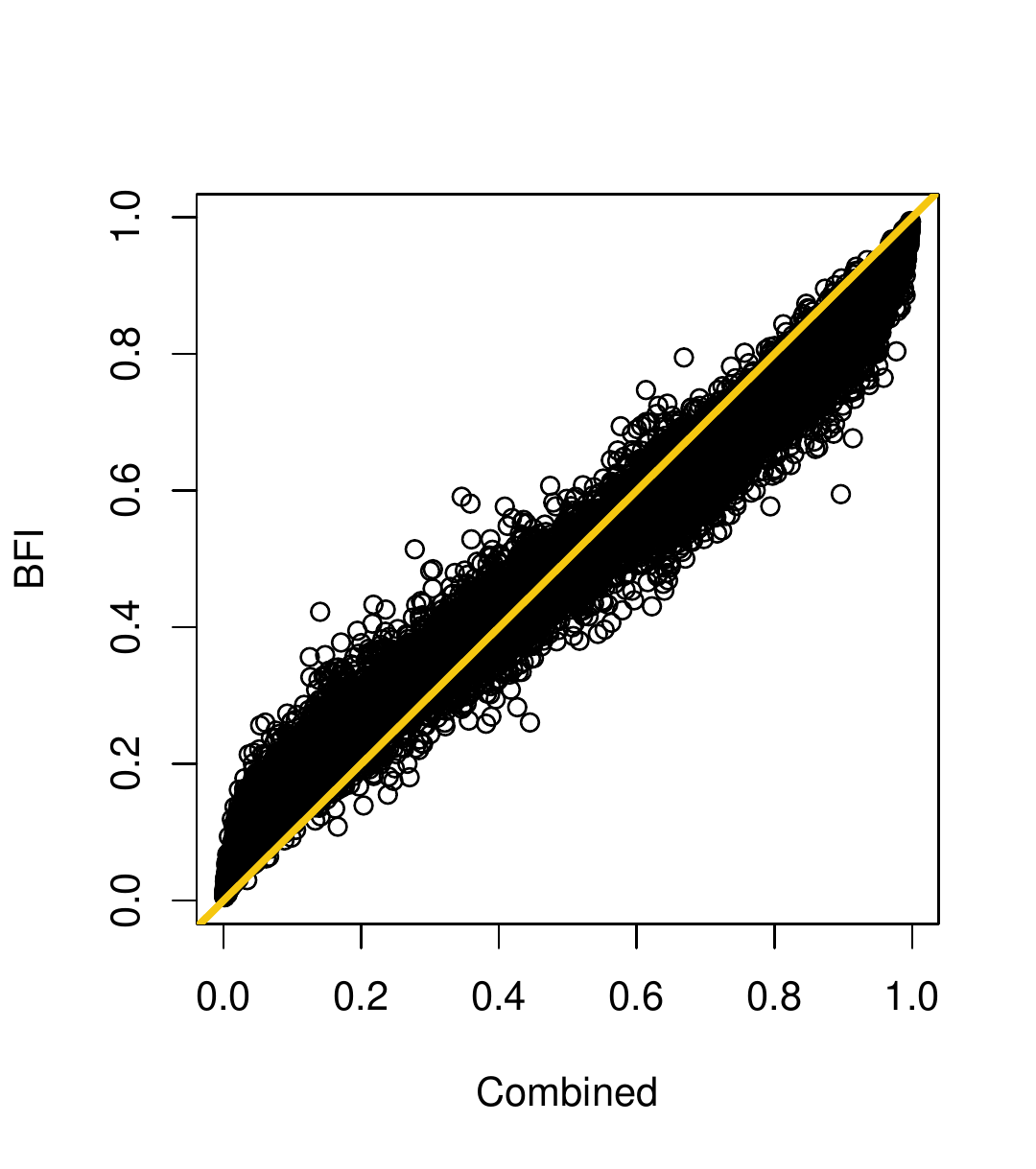}
\includegraphics[width=5.9cm,height=7cm]{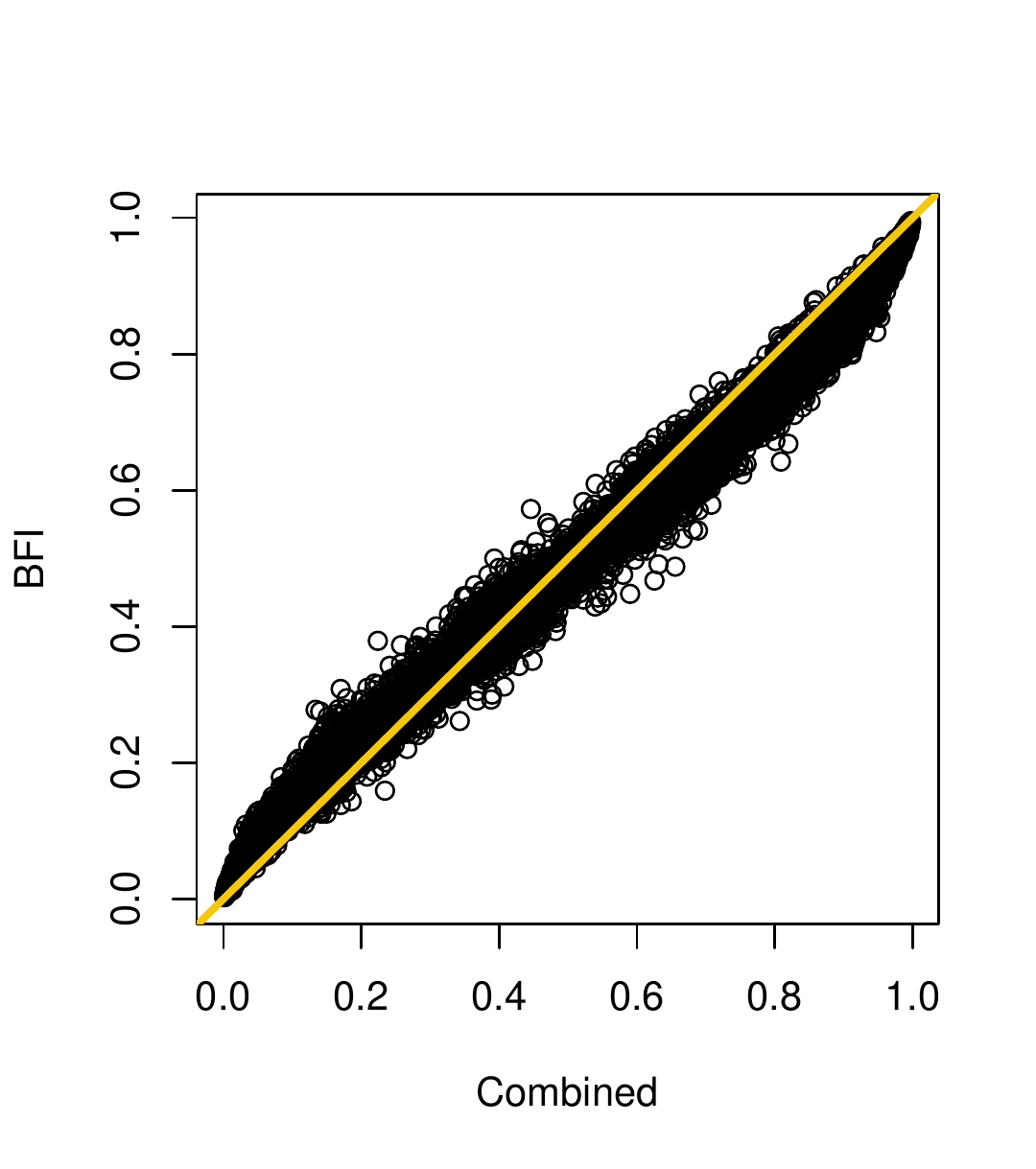}
\includegraphics[width=5.9cm,height=7cm]{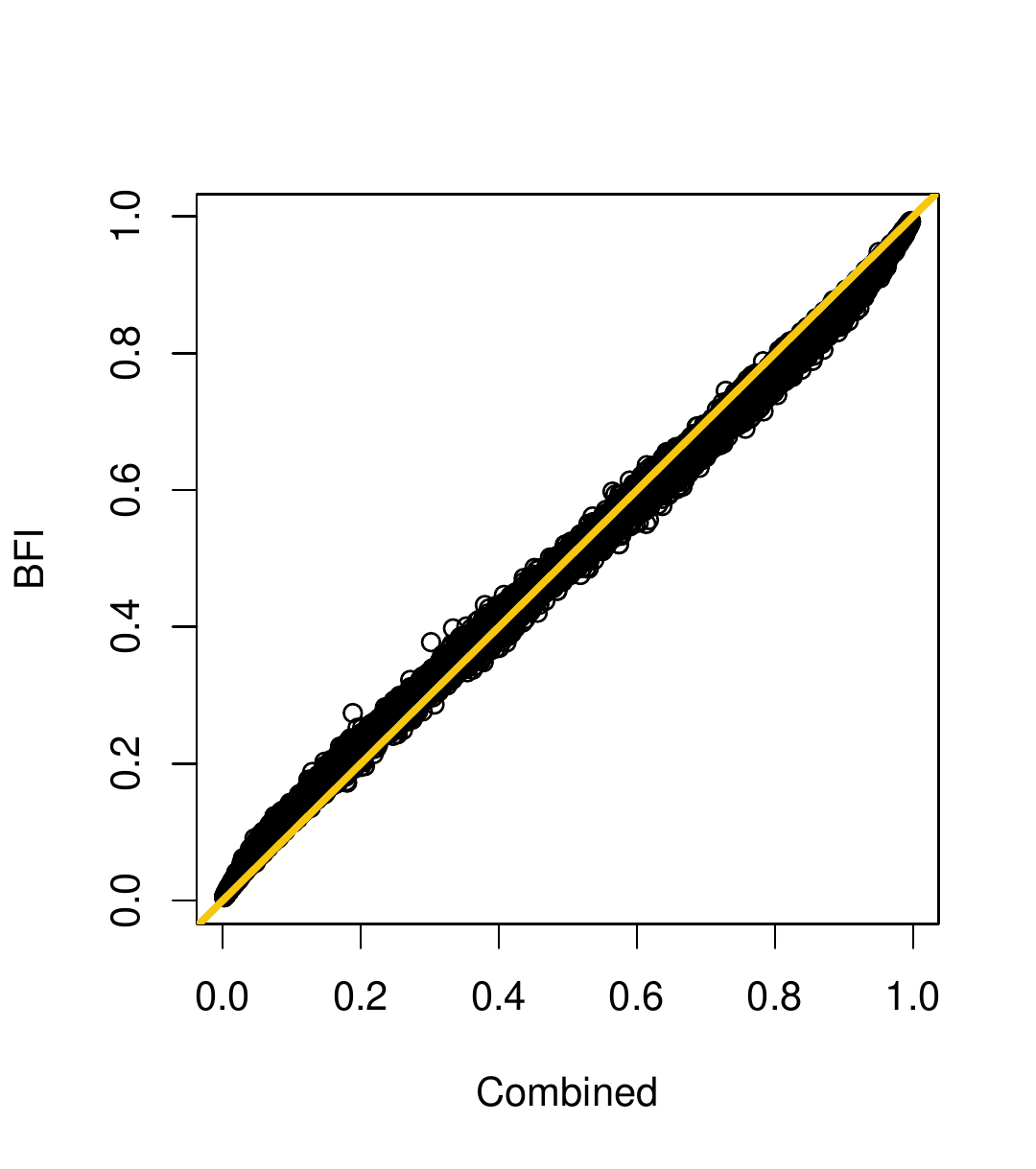}
\caption{
Prediction probabilities based on the estimates $\widehat{\bbeta}_{b}$ (BFI protocol, vertical axis) versus those based on $\widehat{\bbeta}_c$ (regression on combined data set, horizontal axis), in the setting with small sample sizes in nine subsets. Here $n_\ell=40$ for $\ell=1\ldots 8$, and $n_9=51$. From left to right: $\lambda=0.01$, $\lambda=0.1$, and $\lambda=1$. The correlation for the three settings are, respectively: $R^2=0.99 , R^2=1.0, R^2=1.0.$.}
\label{fig:homosmall}
\end{center}
\end{figure}

If the sample sizes of the subsets are small compared to the number of covariates in the model, there is a risk of overfitting. Our present data set consists of three data subsets, ranging in size from 49 to 216 patients. Especially the subset with 49 patients is small, and overfitting can be addressed by taking values for $\lambda$ that are not (too) close to zero. In this subsection we consider a more extreme situation in which multiple subsets are small. We create such data subsets from our original hospital data by randomly dividing all patients over eight subsets of size 40 each, plus one additional subset of 51 patients. The combined data set $\bD$ remains the same, hence the same parameter estimates are found based on the combined data set. We perform the simulation study as described in Subsection \ref{sub:homo} to study the effect of modest subset sample sizes on the estimation and prediction agreement. We do this for different values of $\lambda$. The values of $MSE_{\beta_k}$ are given in Table \ref{Table:MSEHomo}. For the patient outcome predictions we found the following MSE values: $MSE_p=2.95 \times 10^{-3}$ for $\lambda=0.01$, $MSE_p=1.08 \times 10^{-3}$ for $\lambda=0.1$ and $MSE_p=0.44\times 10^{-3}$ for $\lambda=1$. Upon increasing the value of $\lambda$, the estimates of the regression parameters are shrunk towards zero and agreement increases. This can be seen by the decreasing values of $MSE_p$, and also from the scatter plots in Figure \ref{fig:homosmall}. For $\lambda=0.01$ (left plot in Figure \ref{fig:homosmall}) the agreement is weaker, and one observes what appears to be overfitting (rotation of  the data cloud relative to the diagonal). For larger values of $\lambda$ the agreement between BFI recombination and regression results on the full data set improves, as expected. As mentioned before, better agreement between the estimates based on the BFI strategy and those found after combining the data does not necessarily mean that the estimates are closer to the true values; we just conclude that the BFI protocol reliably predicts the  results of regression on the integrated data set.   

We repeated the simulation study for $\lambda=0.01$ in the combined data set and $\lambda=0.1$ or $\lambda=1$ in the subsets. The results are very similar as before.  

\subsection{Heterogeneity across populations}
\label{sub:diff}

\noindent
{\bf{Different covariate distributions and effects}}\\
In this subsection we no longer use randomizations, but return to the original data to study agreement between BFI and direct regression on the combined set in terms of parameter estimates and outcome predictions, in the setting where (as is the case in our original data) the populations from the different hospital types are expected to differ. The combined set priors and subset priors were chosen to be identical, Gaussian with zero average and $\bLambda=\bLambda_\ell$ for all $\ell$, with $\bLambda$ a diagonal matrix with $\lambda$ on the diagonal.    
The parameter estimates obtained via the BFI protocol and those found via regression on the combined data set are given in the Table \ref{tab:local}. From this table we conclude that, even if there are differences in the covariate effects between the populations where the subsets where sampled from,  the BFI-estimated regression parameters are still very similar to those that would have been found upon first combining all data in a single set. For the comparisons of the predictions, as before in every cycle 10\% of the individuals are selected randomly for prediction, and not used for estimation. We quantify the agreement in patient outcome probabilities again by (\ref{exp:MSEp}). For $\lambda=0.1$ we found $MSE_p=0.36\times 10^{-3}$. A scatter plot for the comparison of the prediction probabilities is shown in Figure \ref{fig:PredProb} (right). It can be seen that the agreement between the predictions is high.\\

%\begin{table}
%\begin{center}
%\begin{tabular}{|l|r|r|r|r|r|}
%\hline
%   & intercept & sex & age & ISS  & GCS \\
%\hline
% peripheral no NSU & $-2.82~~ (0.94)$ & $-1.74~~ (0.90)$ &  $1.54~~ (0.56)$ &  $1.90~~ (0.70)$ & $-2.06~~ (0.65)$ \\
% peripheral with NSU & $-0.97~~ (0.33)$ & $-0.34~~ (0.31)$ & $1.90~~ (0.45)$ &  $0.52~~ (0.36)$ & $-1.82~~ (0.40)$\\
% academic hospitals & $-2.22~~ (0.36)$ &  $0.07~~ (0.25)$ &  $1.27~~ (0.29)$ &  $0.30~~ (0.26)$ & $-2.36~~ (0.40)$\\
% \hline
% Combined ($\bbeta_c$) & $-1.70~~ (0.22)$ & $-0.15~~ (0.18)$ &  $1.36~~ (0.20)$ & $0.55~~ (0.19)$ & $-1.98~~ (0.24)$\\
% BFI ($\bbeta_b$) & $-1.51~~ (0.22)$ & $-0.12~~ (0.18)$ & $1.23~~ (0.21)$ & $0.51~~ (0.19)$& $-1.74~~ (0.24)$\\
%\hline
%\end{tabular}
%\caption{
%Estimates of the regression parameters $\bbeta$ in each center (including estimated standard deviations), after combining the data (also given in Table \ref{tab:local}, given here for comparison), and based on the BFI strategy. Here $\lambda=0.1$. } 
%\label{tab:MSEhetero}
%\end{center}
%\end{table}

\medskip

\noindent
{\bf{Different prevalence}}\\
The mortality rates in the three subsets differ (see Table \ref{Tab:data} in Appendix A). This may not be explained completely by the covariates in the model. The prevalence of mortality is reflected in the intercepts of the logistic regression models for the three subsets. To address this observed mortality difference we allow the intercept to vary across hospital types, and we aggregate the estimates from the different subsets as explained in Subsection \ref{subsec:nuisance}. This yields a single logistic regression model that includes a hospital type specific intercept, but no general intercept. For comparison, we also fitted a logistic regression model with hospital type intercepts based on the combined data (instead of a model with a general intercept and a reference hospital type). The combined set priors and subset priors were chosen to be identical, Gaussian with zero mean and $\bLambda=\bLambda_\ell$ for all $\ell$, and $\bLambda$ a diagonal matrix with $\lambda$ on the diagonal.    
For $\lambda=0.1$ and $\lambda=1$ the results are shown in Table \ref{tab:diffprev}. We see very satisfactory agreement between the estimates. We also see, as expected, that the estimates $\widehat{\bbeta}_c$ are typically shrunk to zero for $\lambda=1$ more than for $\lambda=0.1$. 
\begin{table}
\begin{center}
\begin{tabular}{|l|l|r|r|r|r|r|r|r|}
\hline
% $\lambda$ &  & intercept & hospital 2 & hospital 3 & sex & age & ISS  & GCS \\
%\hline\hline
%$\lambda=0.1$ &  BFI &  -1.500 & 0.331 & -0.463 & -0.051 & 1.318 & 0.519 & -1.816 \\
% & Combined & -1.606 & 0.455 & -0.480 & -0.161 &  1.399 &  0.570 & -1.956        \\
%\hline
%$\lambda=1$ &  BFI & -1.659 & 0.510 & -0.404 & -0.024 & 1.290 & 0.568 & -1.767 \\
% & Combined &  -1.342 &  0.239 & -0.612 & -0.136 & 1.279 &  0.551 & -1.816       \\
%\hline
%\hline
 $\lambda$ &  & intercept 1 & intercept 2 & intercept 3 & sex & age & ISS  & GCS \\
\hline\hline
$0.1$ &  BFI  &   $-1.45~(0.52)$ & $-1.07~(0.31)$ & $-1.94~(0.30)$ & $-0.12~(0.18)$ & $1.30~(0.21)$ &  $0.53~(0.19)$ & $-1.82~(0.24)$ \\
 & Comb  & $-1.60~(0.47)$ & $-1.13~(0.32)$ & $-2.07~(0.29)$ & $-0.16~(0.18)$ &  $1.39~(0.21)$ &  $0.57~(0.19)$ & $-1.95~(0.24)$        \\
\hline
$1.0$ &  BFI &  $-1.17~(0.39)$ & $-0.94~(0.27)$ & $-1.79~(0.25)$ & $-0.13~(0.17)$ & $1.20~(0.18)$ &  $0.54~(0.17)$ & $-1.69~(0.20)$ \\
 & Comb  &  $-1.26~(0.41)$ &  $-0.98~(0.29)$ & $-1.85~(0.25)$ & $-0.13~(0.17)$ & $1.25~(0.19)$ &  $0.54~(0.18)$ & $-1.76~(0.20)$       \\
\hline
\end{tabular}
\caption{
Estimates of the regression coefficients (including estimated standard deviations) computed with the BFI strategy ($\hat\bbeta_b$), and those obtained after first combining the data ($\hat\bbeta_c$), when taking into account the differences in mortality prevalence in the three hospital types by including subset specific intercepts. }
\label{tab:diffprev}
\end{center}
\end{table}

\section{Discussion}
\label{sec:Discussion}
In this paper we have discussed a one-shot BFI strategy for estimating parameters in parametric models, where the data in multiple centers can not be combined for analysis. This is for instance of interest in medical settings in which large central registries do not exist and the data sets in individual medical centers or hospitals cannot be easily combined, because of e.g., privacy legislation. A major advantage of the proposed BFI strategy is that, in contrast to most FL strategies, not only parameter estimates but also the uncertainties of these estimates are computed systematically. These uncertainties reflect the differences and uncertainties in the individual centers. In medicine, complementing parameter estimates with error bars is essential in order to interpret correctly the estimation results.  

In this study we  have performed numerical simulations in multiple settings, with real medical data in which the patient outcomes were binary, and with realistic sample sizes. They reveal a very good agreement between the parameter estimates and patient outcome predictions obtained with the BFI procedure, with those  found after first combining all data in a single integrated set.  
We conclude that for the given data and the chosen regression model (logistic) the inference results based on the combined data set can be computed reliably {\it a posteriori} from the inference results on the local centers; hardly any information is lost by not being able to combine the data sets.   

In the proposed BFI strategy we  approximated the posterior parameter distribution around the MAP estimator $\widehat{\btheta}$ by a Gaussian distribution with mean $\widehat{\btheta}$ and a covariance matrix equal to minus the inverse curvature matrix in $\widehat{\btheta}$. For large sample sizes, the MAP estimator will approach the maximum likelihood estimator and minus the inverse curvature matrix will approach the Fisher information matrix (Bernstein-von Mises theorem\cite{Vaart}). If the sample sizes in the centers are small (compared to the number of covariates or the dimensionality of parameter space) these identification may no longer hold. In particular, if the sample size is close to or smaller than the number of covariates, the ML estimator will be inaccurate or not even defined, whereas the MAP estimator is still well defined by the presence of the ridge penalty in the likelihood (for a Gaussian prior). However, from the calculations and the simulation studies it is not clear whether in this situation the Gaussian approximation is sufficiently accurate; i.e., whether the second order Taylor expansion of the logarithm of posteriori density is accurate. A higher order approximation (third or higher order) may give better results, but will change the derivation of the estimators $\widehat{\btheta}$ and $\widehat{\bA}$. This will be part of a new project in which we aim to consider high dimensional models.

If, in one of the local centers, the posterior density does not have a unique maximum, the MAP estimator is not uniquely defined in this center and the BFI methodology cannot be used. This problem may only arise if the sample size is small. Choosing a different matrix $\bLambda$ may solve the problem of the unique maximum. If the log posterior density is multimodal (has local maxima), the theory still holds, but one has to be careful when  numerically optimizing the log posterior density. 

In the paper we have considered two priors: the Gaussian prior and the flat noninformative prior (which in principle equals the Gaussian prior with infinite variance). These priors are not randomly chosen. The form of these priors yields a quadratic expression in the parameters in the equation (\ref{eq:log-equality}), which is essential in the derivation of the BFI estimators. Other prior distributions can be used as well as long as they give this quadratic expression. The (inverse) covariance matrix of the Gaussian prior can be chosen by the researcher. The choice may affect the degree of regularization.

While in most of our simulations we have for simplicity used the same inverse covariance matrices for the data subsets and the combined data set, the derivations in sections 2 and 3 are in fact more general and allow these matrices to be distinct. This freedom offered by the theory to vary the covariances of the priors could for instance be employed to regularize/penalize smaller data subsets more than larger ones as was shown in subsections \ref{sub:homo} and \ref{sub:small}. Also if data in some centers are more reliable or ``cleaner'' than in other centers, or if the populations in some centers seem to be more representative for the population of interest (e.g., if the inclusion criteria differ across centers), different prior covariances can be used. By allowing this generalization in the theory and the software package it is up to the user to assume equal priors or not.

The main aim of the simulation studies and the data analysis (Section \ref{sec:simstudies}) was to illustrate the proposed methodology. The motivation of this research comes from oncology. Some cancers are extremely rare and only small data sets are available. This makes it impossible to study overall survival as the numbers of events per data-set are low. 
In this paper we have developed and applied the BFI methodology for parametric (regression) models, and in particular for GLMs. The BFI methodology is also applicable in more complex models and settings, such as models for survival data that are subject to censoring. %, or random effects models. 
In particular, the development of BFI protocols for semi-parametric models, like the Cox proportional hazards model\cite{Cox} which is widely used in medical data analysis, is more complex, because of its infinite dimensional parameter space. This will be our next project: to develop the BFI methodology for cancer data with time-to-event outcomes. 

In conclusion, we have shown that by harnessing systematically and accurately the cumulative power of multiple disjunct data sets without actually combining these data sets, the BFI methodology can reduce significantly the sizes of data sets required for extracting reliably statistically significant predictive or prognostic patterns (if such patterns are present). While with conventional methods only very strong regularities and associations can be identified in small data sets (small compared to the dimension of the model), with the new approaches more subtle ones may become detectable.  \\

\bigskip

\noindent
{\bf{Acknowledgements}}\\
We thank dr.\ J. Draaisma for making his data available. We also like to thank two reviewers and the associate editor for their valuable comments.

\bigskip

\noindent
{\bf{Software}}\\
The R package {\it BFI} and a manual are available at github: {\tt 
 https://hassanpazira.github.io/BFI} . A more detailed manual can be found at {\verb| https://github.com/hassanpazira/BFI/ | }.

\bigskip

\noindent
{\bf{Data availability statement}}\\
The trauma data are available in the R package BFI. 

\bigskip

\noindent
{\bf{Funding}}\\
This research was supported by an unrestricted grant of Stichting Hanarth Fonds, The Netherlands. 

%\bibliographystyle{apalike} %apalike %plain
%\bibliographystyle{natbib} %apalike %plain
%\bibliography{Bibfor-BFI-GLM} 

%\end{document}

\section*{Appendix A: Details on the data}
The data originate from multiple hospitals which can be categorised in three groups as: peripheral hospitals without a neuro-surgical unit,  peripheral hospitals with a neuro-surgical unit, and academic medical centers. They were collected  between October 1984 and October 1985, in 12 hospitals from one of the three categories above. The original aim in collecting these data was to compare hospitals from the three categories, in terms of differences in rates of management error \cite{Draaisma}. In the data in their present form (that have since been made available for educational purposes) there is no longer any reference to the individual hospitals; there is only a variable indicating in which of the three hospital categories the data were recorded. In the absence of more detailed information we assume, for simplicity and in the terminology of the present study, that there are only three data subsets or centers, which correspond to three categories. In this paper the data are used only to study the performance of the BFI methodology. 

An overview of the population characteristics is given in Table \ref{Tab:data}.

\begin{table}[h]
\begin{center}
\begin{tabular}{|l|r|r|r|r|r|r|}
\hline
Data subsets & number   &  mortality &      age &       sex  & ISS     & GCS  \\
              &   $n_\ell$    &  \%      & median & \% females & median & median \\
\hline
peripheral without NSU   & $49$  & $43$ & $30$ & $22$ & $41$ & $10$  \\
peripheral with NSU & $106$ & $40$ & $29$ & $24$ & $33$ & $10$  \\
academic hospitals        & $216$ & $22$ & $31$ & $30$ & $29$ & $14$  \\
\hline
combined data            & $371$ & $30$ & $29$ & $27$ & $30$ & $12$  \\  
\hline
\end{tabular}
\end{center}
\caption{Data characteristics in the three subsets (here indicating different hospital types) and in the combined data set. NSU stands for `neuro-surgical unit'. }
\label{Tab:data}
\end{table}

\end{document}